\documentclass[]{aa}
\usepackage{txfonts}
\usepackage{graphicx}
\usepackage{natbib}
\usepackage{subfigure}
\usepackage{ulem}
\usepackage{wasysym}

\def \LA {LA }
\def \Msun {M_{\odot}}
\def \RHK {R'_\mathrm{HK}}

\begin{document}
\title{The Hercules-Lyra Association revisited\thanks{Based on observations made with ESO Telescopes at the Paranal Observatory under programs ID: 380.C-0248(A) (Service Mode, VLT-Yepun) and ID: 074.C-0084(B) (on 2005 Jan 06, VLT-Yepun).}\thanks{Based on observations collected at the Centro Astron\'omico Hispano Alem\'an (CAHA) at Calar Alto, operated jointly by the Max-Planck Institut f\"ur Astronomie and the Instituto de Astrof\'isica de Andaluc\'ia (CSIC).}}
\subtitle{New age estimation and multiplicity study}
\author{T. Eisenbeiss\inst{\ref{aiu}}
\and M. Ammler-von Eiff\inst{\ref{TLS},\ref{IAG},\ref{MPS}}
\and T. Roell\inst{\ref{aiu}}
\and M. Mugrauer\inst{\ref{aiu}}
\and Ch. Adam\inst{\ref{aiu}}
\and R. Neuh\"auser\inst{\ref{aiu}}
\and T.~O.~B.~Schmidt\inst{\ref{aiu}}
\and A. Bedalov\inst{\ref{aiu},\ref{LUB},\ref{zag}}}
\institute{Astrophysical Institute and University Observatory, Friedrich-Schiller Universit\"at, Schillerg\"asschen 2-3, 07745 Jena, Germany, \email{eisen@astro.uni-jena.de} \label{aiu}
\and
Th\"uringer Landessternwarte, Sternwarte 5, 07778 Tautenburg, Germany\label{TLS}
\and
Georg-August-Universit\"at, Institute for Astrophysics, Friedrich-Hund-Platz 1, 37077 G\"ottingen, Germany\label{IAG}
\and
Max Planck Institute for Solar System Research, Max-Planck-Strasse 2, 37191 Katlenburg-Lindau, Germany\label{MPS}
\and
Department of Physics and Astronomy K.U.Leuven, Celestijnenlaan 200D, B-3001 Leuven\label{LUB}
\and
Faculty of Natural Sciences, University of Split, Teslina 12. 21000 Split, Croatia\label{zag}
}

\date{}

\abstract
{
The Hercules-Lyra association, a purported nearby young moving group, contains a few tens of {zero age} main sequence
stars of spectral types F to M . The existence and the properties of the Her-Lyr association are controversial and discussed in the literature.
} 
{The present work reassesses the properties and the member list of the Her-Lyr association based on kinematics and age indicators. Many objects form multiple systems or have low-mass companions and so we need to properly account for multiplicity.
}
{
We use our own new imaging observations and archival data to identify multiple systems. The colors and magnitudes of kinematic candidates are compared to isochrones. We derive further information on the age based on Li depletion, rotation, and coronal and chromospheric activity. A set of canonical members is identified to infer mean properties. Membership criteria are derived from the mean properties and used to discard non-members.
}
{
The candidates selected from the literature belong to 35 stellar systems, $42.9\,$\% of which are multiple. Four multiple systems (V538 Aur, DX Leo, V382 Ser, and HH Leo) are confirmed in this work by common proper motion. An orbital solution is presented for the binary system which forms a hierarchical triple with \object{HH\,Leo}.

Indeed, a group of candidates displays signatures of youth. Seven canonical members are identified 
The distribution of Li equivalent widths of canonical Her-Lyr members is spread widely and is similar to that of the Pleiades and the UMa group. Gyrochronology gives an age of $257\pm46$\,Myr which is roughly in between the ages of the Pleiades and the { Ursa Major} group. The measures of chromospheric and coronal activity support the young age.

Four membership criteria are presented based on kinematics, lithium equivalent width, chromospheric activity, and gyrochronological age. In total, eleven stars are identified as certain members { including co-moving objects} plus additional 23 possible members while 14 candidates are doubtful or can be rejected. A comparison to the mass function (MF), however, indicates the presence of a large number of additional low-mass members, which remain unidentified.
}


\maketitle

\section{Introduction}

{ The Hercules-Lyra (Her-Lyr) association is one of the closest young moving groups (MGs). The members of moving groups, also called stellar kinematic groups, are identified by similar space motion. In many cases, common space motion is related to spatial clustering, for example in the well-known cases of the Hyades and Ursa Major (UMa, or Sirius) moving groups and the eponymous open clusters. The discovery of additional distant and co-moving stars also led to the concept of superclusters \citep{1994gsso.conf..191E}. The third large nearby group next to the Hyades and UMa groups is the local association (LA) that is associated with the Pleiades. Its members are young with a wide range of ages of $100$\,Myr and younger. In contrast to the Hyades and UMa, the LA clearly lacks aspects of spatial coherence and embeds clusters and associations like $\alpha\,$Per, IC\,2602, and Sco-Cen \citep{2001MNRAS.328...45M, 2004ARA&A..42..685Z}. In the past decade, many nearby young moving groups have been identified with ages below $100$\,Myr \citep{2004ARA&A..42..685Z, 2008hsf2.book..757T}. They are located at average distances of about $30\,$\,pc and beyond \citep{2008hsf2.book..757T}.}


{ At even closer distances}, young stars similar to members of the LA 
and the Pleiades were noticed earlier \citep{1987ApJ...317..787Y,1987AJ.....93..920S}, while {it was not before \citet{1998PASP..110.1259G}  that these stars were considered a close moving group}. This group was previously unnoticed because of the lack of bright stars unlike the UMa group, and because of its spatial scatter due to its proximity \citep{2004AN....325....3F}.

\begin{figure*} 
\centering
\includegraphics[width=17cm,clip=true,trim=0 0 0 0]{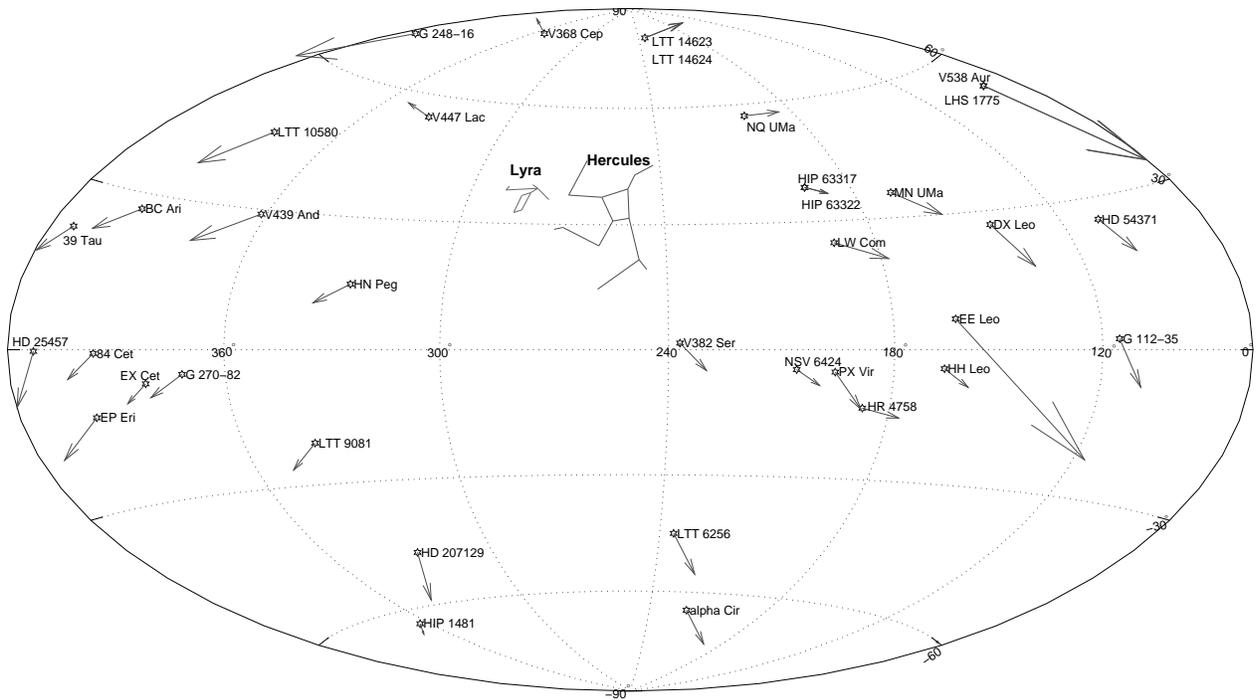} 
\caption{All Her-Lyr membership candidates and their locations on the sky. 
The arrows indicate the proper motion.  The names of the stars are indicated.}
\label{hap}
\end{figure*}

{ In more detail}, \citet{1998PASP..110.1259G} searched the Hipparcos catalog for young solar analogues within 25 pc of the Sun. He noticed two distinct groups in kinematic space, one of which was identified with the UMa group. { Another group of five stars, V439 And, DX Leo, MN UMa, HN Peg, and NQ UMa,} displays kinematics similar, but not equal, to the \LA with the radiant being located in the constellation Hercules (Fig.~\ref{hap}).

In his volume-complete study of northern G- and K-type stars in the solar neighborhood within 25\,pc, \citet{2004AN....325....3F} detected a concentration of stars with higher rotation rates typical of stellar youth coinciding in $UV$ space with the Hercules moving group { identified by} \citet{1998PASP..110.1259G}. Depending on the details of the selection of up to 15 candidates, the radiant may also be located in the constellation Lyra, thus the name Her-Lyr association \citep{2004AN....325....3F}. The equivalent widths of the H$\alpha$ and lithium lines of many of these stars resemble those of UMa group members for which \citet{2004AN....325....3F} gives an age of approximately $200\,$Myr.

While \citet{2004AN....325....3F} founded the { very} existence of the group on  the statistically enhanced concentration in kinematical space, \citet{2006ApJ...643.1160L} gave additional quantitative criteria for Her-Lyr membership. From the sample of \citet{2001MNRAS.328...45M} of late-type members of young stellar kinematic groups, they constructed an initial sample adding twelve stars to Fuhrmann's sample. A velocity criterion was derived taking the average $UV$ velocity of Fuhrmann's sample and adopting the dispersion of 6 km\,s$^{-1}$ of the Castor moving group as the tolerance limit. In addition, the lithium equivalent width and the position in the color-magnitude diagram were required to be compatible with a young age of 200 Myr.

\citet{2010A&A...521A..12M} also investigate the properties of young nearby stars and put the young nearby MGs into context, noticing that "\ldots { the young MGs (8-50\,Myr) are probably the most immediate dissipation products of the youngest associations.}" 
In their conclusion they find that "\ldots {All previously proposed members of AB\,Dor or Her-Lyr fall into our classification of probable LA members, \ldots}".
The Her-Lyr group might therefore be a sub-group of the \LA or even indistinguishable from it.

The aim of this article, however, is not to discuss the { very} existence of the moving group. Its emergence in the $UV$-plane as a concentration of fast rotating young stars was well established by \citet{2004AN....325....3F}. In Fig. \ref{hap} all Her-Lyr candidates discussed in this paper are displayed along with their proper motion in a Hammer-Aitoff projection. The radiant coincides with the constellations Hercules and Lyra. No kinematic candidates are located close to the radiant or the point of convergence. Any kinematic candidates in these areas would have small, if not negligible, proper motion and { a value} of radial velocity which is close to the total space motion of the Her-Lyr association. { Therefore,} the lack of detections in these regions of the sky does not necessarily mean a real lack of Her-Lyr candidates but may instead reflect the presence of an { observational selection effect}. The Her-Lyr association in this area will have very small proper motions, moving mainly in the radial directions and may not be identified in the initial kinematic sample.

In the present work we want to establish the mean properties of, as well as a refined member list of, the Her-Lyr association. In addition to the tools used by \citet{2004AN....325....3F} and \citet{2006ApJ...643.1160L} we investigate the multiplicity of the Her-Lyr candidates in Sect.~\ref{multi}, revealing the low-mass membership candidates, typically well-known sources that are visually resolved because of the proximity of the parent stars.
Finding late-type companions to the stars may indeed give better constraints on ages derived via isochrone-fitting (Sect.~\ref{photo}) as late-type stars arrive on the main sequence later. 

The assessment of an age has been unsure in the past as it is based on a few stars only, assuming similarity to the UMa group and adopting its age estimate \citep{2004AN....325....3F}. In addition to the study of the lithium equivalent width (Sect.~\ref{lithium}) we use gyrochronology \citep{2007ApJ...669.1167B,2009IAUS..258..345B,2009IAUS..258..375M} in Sect.~\ref{gyro} as a tool to investigate the age of the membership candidates. In addition we analyze the chromospheric activity index $(R'_{\mathrm{HK}})$
 in Sect.~\ref{calcium} and the coronal activity (X-ray luminosity) in Sect.~\ref{xrays}.

Another open question is the rather arbitrary adoption of the Castor velocity dispersion for the kinematic criterion given by \citet{2006ApJ...643.1160L}. Interestingly the dispersion of their final sample is lower.

To define the properties and the list of members a new approach is followed in Sect.~\ref{kine} in order to avoid any assumptions or { selection effects}, starting from the concentration of rapid rotators in the volume-complete work of \citet{2004AN....325....3F}. The youth indicators of these stars will be studied as a whole, i.e., the distribution in lithium equivalent width and ({gyrochronological}) age. If a young Her-Lyr association exists as its own entity, a group of young stars will stand out in all of the properties studied, even if intrinsic variations might be present. It is assumed that this group of stars is composed of the canonical members which define the overall properties of the Her-Lyr association.
In Sect. \ref{indstar} individual interesting stars among the Her-Lyr candidates are addressed with emphasis given on multiplicity. 

Based on this information we present and discuss an updated list of Her-Lyr members and non-members in Sect.~\ref{concl}. We evaluate the mass function of the membership candidates before we conclude in Sect. \ref{outlook}.


\section{Definition of the Sample}
{ We started with the Her-Lyr candidates defined by \citet{2004AN....325....3F}, with the four stars of the Hercules moving group \citep{1998PASP..110.1259G} already included. We added new candidates introduced by \citet{2006ApJ...643.1160L} and \citet{2008MNRAS.384..173F}. We never removed a candidate beforehand, even if it was later rejected for some reason. Also we did not add candidates, which had been previously excluded within the same study in which they where considered as candidates for the first time. In particular we did not include objects, classified as members of the \LA$\!\!$ if they had never been taken into account as Her-Lyr candidates before. Our multiplicity study, which is described in detail in the following section, increased the number of candidates again, but did not increase the number of stellar systems. Most of these companions had been discovered previously. An intense study of the literature brought up additional candidates, including spectroscopic and sub-stellar companions.
}

\section{Multiplicity}
\label{multi}
{ Our multiplicity study focuses on archival work and the search for common proper motion pairs. Using archived surveys along with state-of-the-art follow-up observations enables the combination of different data sources  
\citep{2010hsa5.conf..379C,2011sca..conf..344A}, offers a way for accurate indirect determinations of quantities like parallax 
\citep{2004ApJS..150..455G} or { the} evolutionary status \citep{2008ApJ...687..566M}, and provides a starting point  for extended surveys (like the Palomar/Keck survey of \citealp{2009ApJS..181...62M}), as well as extreme findings (like the wide ultra-cool binary by \citealt{2007A&A...462L..61C}).}

Within our ongoing multiplicity study and based on the \citet{2004AN....325....3F} and \citet{2006ApJ...643.1160L} sample we initiated an archival search for faint companions.
Given that most of the Her-Lyr candidates are within $25$\,pc of the Sun and hence bright, we can use archival data, for example,  the Two Micron All Sky Survey (2MASS) \citep{2003tmc..book.....C}. Furthermore, we used Schmidt plates from the Palomar and ESO sky surveys, scanned by the SuperCOSMOS machine \citep{1998MNRAS.298..897H}, providing best available astrometric accuracy. { We found co-moving objects for six nearby stars: \object{V538\,Aur}, \object{DX\,Leo}, \object{HH\,Leo}, \object{HIP 63317}, \object{LTT14623}, and \object{V382Ser}. Each one of the companions except for one were known already, so we published the new one in \citet{2007AN....328..521E}}. This section gives an overview of the multiple systems among the Her-Lyr stars included in this ongoing study.


\begin{figure*}
\centering
\subfigure[\object{V538\,Aur} = \object{HD\,37394} and \object{LHS\,1775}]{\includegraphics[width=6cm]{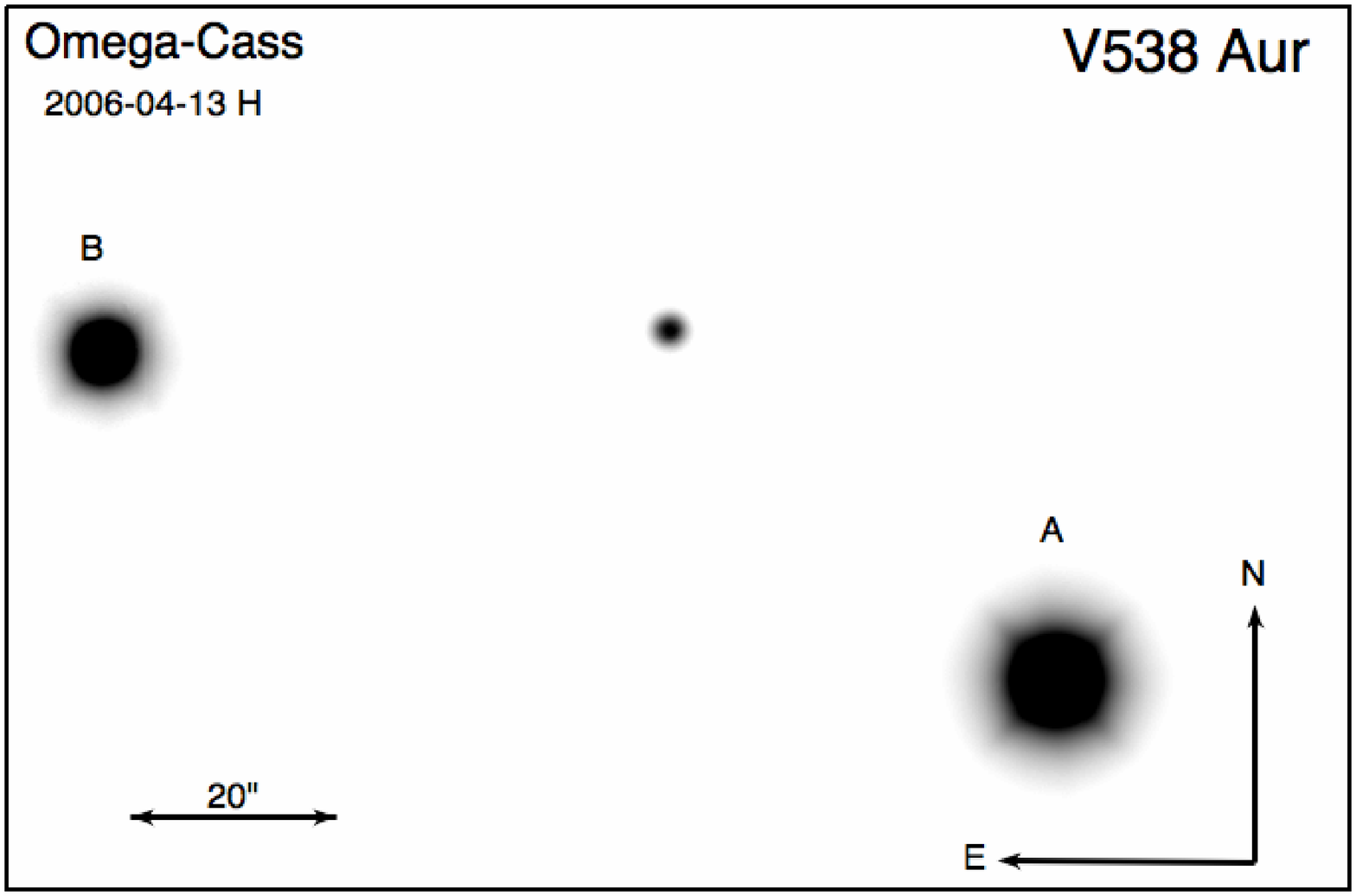}}
\subfigure[\object{DX\,Leo} = \object{HD\,82443} and \object{GJ\,354.1\,B}]{\includegraphics[width=6cm]{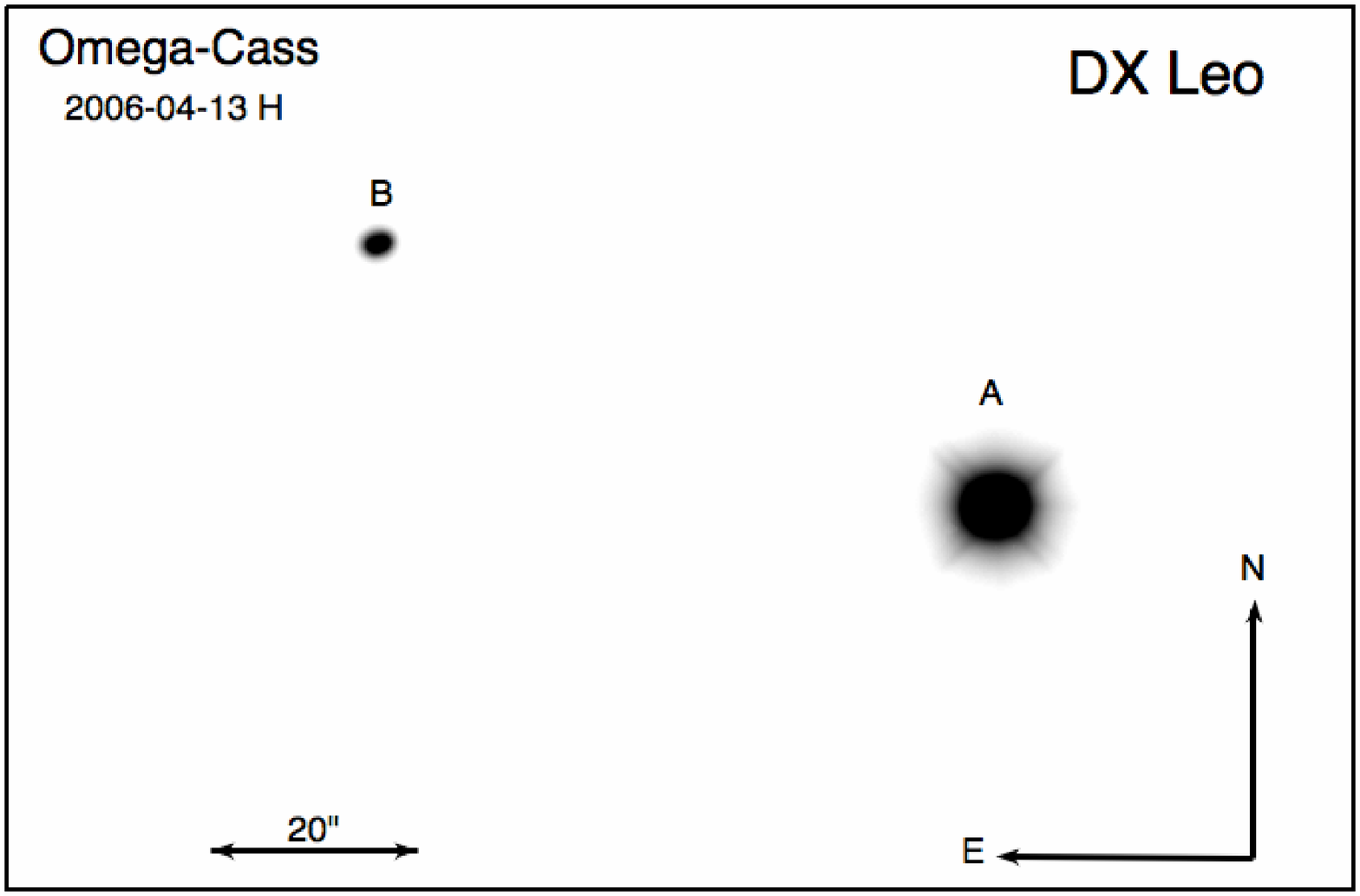}}
\subfigure[\object{V382\,Ser} = \object{HD\,141272} and \object{HD\,141272\,B}]{\includegraphics[width=6cm]{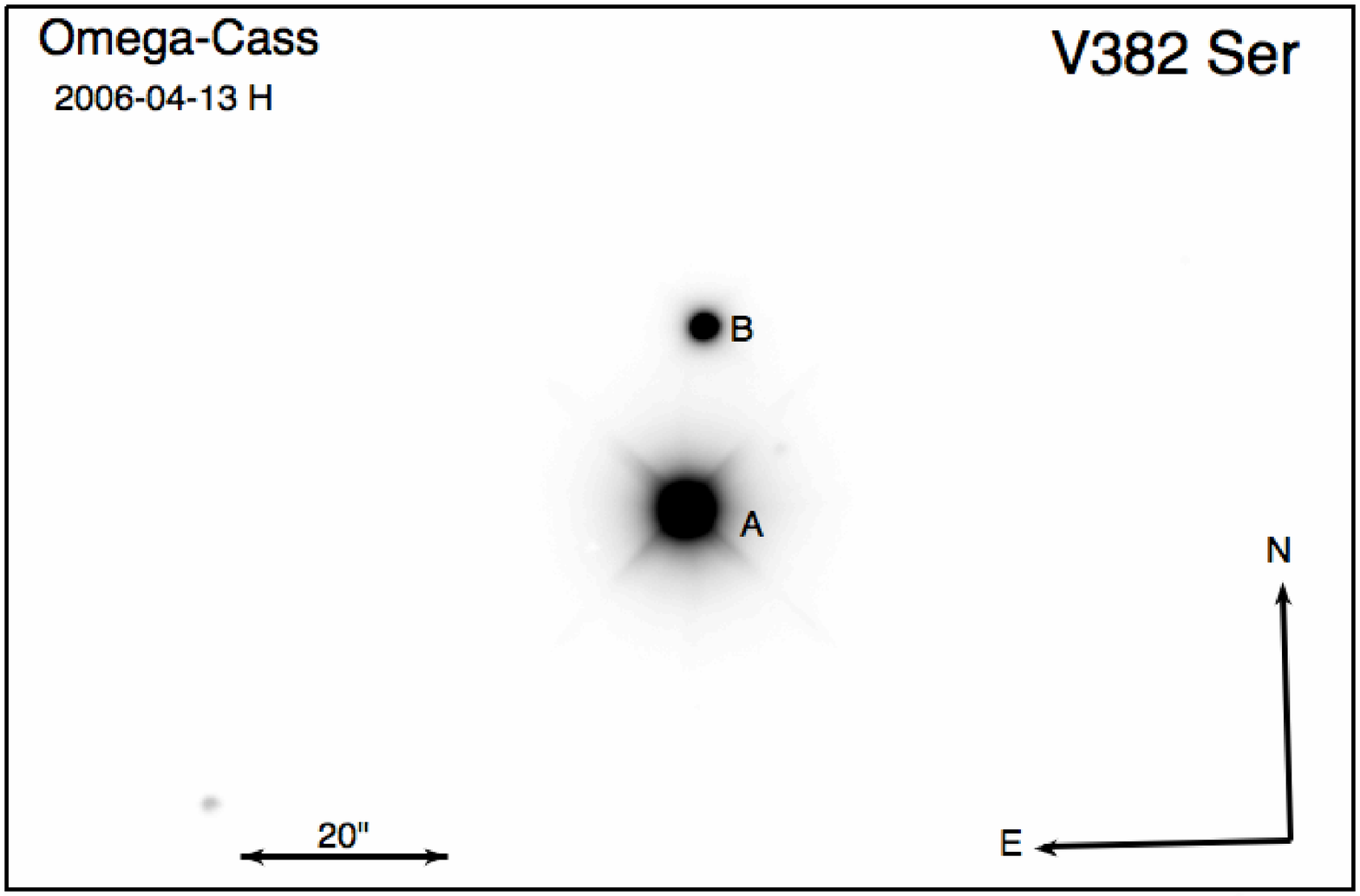}}
\caption{Three Her-Lyr candidates (labeled "A") and their visual companions (labeled "B"), observed with the Calar-Alto $3.5$\,m telescope and $\Omega$-Cass. \object{V382\,Ser} was already published in \citet{2007AN....328..521E}.}
\label{visbin}
\end{figure*}

The analysis follows the procedures outlined in \citet{2007AN....328..521E}. Taking the scanned Schmidt plates as first epoch and 2MASS as the last, we identify wide stellar binaries among Her-Lyr members as common proper motion pairs. Calibration and uncertainties are derived in a statistical way using the (non-moving) background stars {in the stellar field observed}. If the Her-Lyr star is saturated in the Schmidt plates, the diffraction spikes are used to determine an accurate position measurement. Because of the large epoch differences, the proper motions derived have uncertainties of only $\sim5\,$mas/yr. {Considering the large proper motion of the nearby Her-Lyr stars and possible companions, this uncertainty corresponds to a relative error of a few percentage points}. Most of the companions we found were already known before but were not taken into consideration by former studies of the Her-Lyr association. 

For some stars we took additional data to look for closer and fainter companions. { The search for common proper motion pairs in the archives is supplemented by own new observations.}
{ We observed with either VLT/NaCo \citep{2003SPIE.4841..944L,2003SPIE.4839..140R} or Calar Alto 3.5m/ALFA
(or Omega-Cass, \citealp{1998SPIE.3354..493L,2000ExA....10...49K}) isolated young nearby stars (with an age of up
to some 100\,Myr and a distance of up to some 100\,pc)
in order to detect possible sub-stellar companions.

Naos-Conica is the adaptive optics imager and spectrograph installed at the Nasmyth B focus of UT4 (Yepun) at ESO's Paranal observatory in Chile. The adaptive optics Naos contains a tip-tilt mirror, as well as a deformable mirror with 185 actuators. The infrared Conica camera is equipped with an Aladdin 3 InSb infrared detector ($1024\times1024$ pixel each with a size of $27\times27\,\mu$ m). The Naco data presented here (Table \ref{obslog}), were taken with Naos's visible wavefront sensor (VIS) using its $14\times14$ lenslet array together with Conica, operated in the HighDynamic mode, using its S13 optics (pixel scale of about 13\,mas/pixel).

Omega-Cass is an infrared camera, which was operated at the Cassegrain focus of the 3.5\,m telescope at the Calar Alto observatory in Spain. The camera is equipped with a Rockwell HAWAII HgCdTe infrared detector ($1024\times1024$ pixel, each with a size of $18.5\times18.5\,\mu$m) operated in the non-destructive readout mode. Omega-Cass could be used either as seeing limited infrared imager or together (as done here) with the adaptive optics Alfa for diffraction limited imaging at the CAHA/3.5m. Alfa is equipped with a tip-tilt mirror, a deformable mirror with 97 actuators, and a visible wavefront sensor. We obtained images of HN Peg and HH Leo with Omega-Cass together with Alfa in the jitter-mode.

Some results have been reported elsewhere (e.g. \citealp{2005A&A...435L..13N}). Here, we present the results on possible Her-Lyr members, see Table \ref{obslog}. { Section \ref{indstar} includes notes on all multiple Her-Lyr candidates with references.}} Figure~\ref{visbin} shows three examples, \object{V538\,Aur} ( = \object{HD\,37394}), \object{DX\,Leo} ( = \object{HD\,82443}), and \object{V382\,Ser} ( = \object{HD\,141272}) and their companions (indicated by the letter B). The images were obtained using the Calar Alto $3.5\,$m telescope and the near-infrared imager $\Omega$-Cass \citep{1998SPIE.3354..493L}. Two more objects are discussed in detail in the following subsections. These data were reduced using {\it ESO Eclipse} \citep{2001ASPC..238..525D} and calibrated using background stars in the field (not visible in Fig.~\ref{visbin} because of the cut levels and the small section of the images shown), listed in the 2MASS catalog. In the worst case, we had five reference stars in the field, still sufficient if field distortion, skew, and other non-linear effects are neglected. 
The reader is referred  to \citet{2007AN....328..521E} for further details. We give the observations log in Table \ref{obslog}. In Table \ref{hlbinca} we give the separation and the position angle with respect to the primaries.

{ The binary status of \object{V382\,Ser} was already published in \citet{2007AN....328..521E}. Because of saturation effects the binary was barely resolved on Schmidt-plates. We succeeded using the Source extractor \citep{1996A&AS..117..393B} and confirmed our results with follow-up observations.}

Furthermore, a multiplicity  study of nearby stars has been carried out by \citet{2010ApJS..190....1R}. 
Nevertheless, the Her-Lyr association was not addressed in particular. 
Combining the results with the present work, the multiplicity census of the Her-Lyr candidates within $25$\,pc should be  complete \citep[see][Sect. 2.2 and 5.2 for details]{2010ApJS..190....1R}\footnote{\citet{2010ApJS..190....1R} combine a large variety of datasets, including data from high angular resolution imaging and high precision spectroscopy. They detected all companions into the planetary regime at all realistic separations.}. 


\begin{table}
\setlength{\tabcolsep}{5pt}
\caption{Observations log}
\label{obslog}
\begin{center}
\scriptsize
\begin{tabular}{llllcc}
\hline\hline
Target	&	MJD	&	Tel./Inst.	&	Filt.	&	scale	&	det. ori.\\
		&[days]	&			&		&	[mas/pix]	&	[$^\circ$]	\\
\hline
V538\,Aur	& 53800	&	CAHA 3.5\,m/$\Omega$-Cass	&H	& $192\pm0.43$	& $358.14\pm0.18$\\
DX\,Leo	& 53800	&	CAHA 3.5\,m/$\Omega$-Cass	&H	& $192\pm0.43$	& $358.14\pm0.18$\\
V382\,Ser	& 53800	&	CAHA 3.5\,m/$\Omega$-Cass	&H	& $192\pm0.43$	& $358.14\pm0.18$\\
HN\,Peg	& 52631	&	CAHA 3.5\,m/ALFA                	&H	& $77.46\pm0.05$	&$18.6\pm0.1$\\
HH\,Leo	& 52631	&	CAHA 3.5\,m/ALFA                	&H	& $77.46\pm0.05$	& $18.6\pm0.1$\\
HH\,Leo	& 53378	&	VLT/NaCo         	&K$\mathrm{_s}$	& $13.19\pm0.02$	& $359.87\pm0.3$\\
HH\,Leo	& 54473	&	VLT/NaCo         	&K$\mathrm{_s}$	& $13.17\pm0.05$	& $359.92\pm0.17$\\
\hline
\end{tabular}
\end{center}
\end{table}
\begin{table}
\caption{Data of known binary pairs taken with $\Omega$-Cass at the Calar--Alto $3.5\,$m telescope. \object{HN\,Peg} was observed with ALFA (Sect. \ref{halphasec}).}
\label{hlbinca}
\begin{center}
\scriptsize
\begin{tabular}{llcr@{.}l@{$\,\pm\,$}r@{.}lr@{.}l@{$\,\pm\,$}r@{.}l}
\hline\hline
Name	&	Companion	&	JD	     	&	\multicolumn{4}{c}{separation}          	&	\multicolumn{4}{c}{position angle}\\
&&[days]&\multicolumn{4}{c}{['']}&\multicolumn{4}{c}{[$^{\circ}$]}\\
\hline
\object{V538\,Aur}	&	\object{LHS\,1775}	&	2453800	&	98&04&0&07	&	71&312&0&181\\
\object{DX\,Leo}	&	\object{GJ\,354.1\,B}	&	2453800	&	65&11&0&10	&	67&236&0&183\\
\object{V382\,Ser}	&	\object{HD\,141272\,B}	&	2453800	&	17&82&0&11	&	352&53&0&39\\
\object{HN\,Peg} &	\object{HN\,Peg\,B}	& 	2452631	&	42&762&0&029&	254&46&0&11\\	
\hline
\end{tabular}
\end{center}
\tablefoot{Position angle measured from north over east.}
\end{table}

\subsection{ALFA observation of \object{HN\,Peg} A \& B}\label{halphasec}
The \object{HN\,Peg} system was observed on 2002 December 23 in the H-band at the 3.5\,m telescope of the Calar Alto observatory in Spain using the adaptive optics system ALFA \citep{2000ExA....10...49K} in combination with the near-IR imager $\Omega$-Cass  (Table \ref{obslog}). The bright primary \object{HN\,Peg}\,A served as natural AO guide star.

\begin{figure}
\centering
\resizebox{0.85\hsize}{!}{\includegraphics{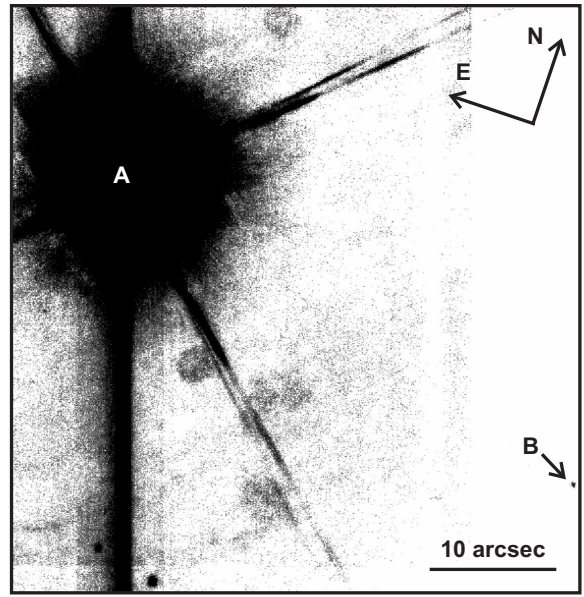}}
\caption{\object{HN\,Peg} A \& B imaged at the $3.5$\,m telescope of the Calar Alto observatory with $\Omega$-Cass and ALFA. \object{HN\,Peg}\,B is almost outside the FoV of the detector, and so it was only imaged at some dither positions.}
\label{HNPeg}
\end{figure}

We used the shortest possible individual integration time of $\Omega$-Cass (0.842\,s) to suppress saturation effects of the IR-detector due to the bright primary star in the observed field of view. Forty-nine of these short exposures were averaged to one image and 24 of these frames were then taken at different telescope positions in jitter-mode for proper sky-background subtraction, which yields 16.5\,min of total integration time on target. For flatfielding, skyflats were taken in the evening and morning twilights. All frames were background-subtracted, flatfielded, and averaged to the image shown in Fig.~\ref{HNPeg}, using the data reduction package \textsl{Eclipse} \citep{2001ASPC..238..525D}. The faint sub-stellar companion \object{HN\,Peg\,B} is detected at the western border of our ALFA image. The same object was detected by \citet{2007ApJ...654..570L} in 2006 June, so our 2002 image is  a 2nd pre-discovery after 2MASS. Relative astrometry, based on the calibration published in \citet{2004A&A...417.1031M}, is summarized in Table~\ref{hlbinca}.

\subsection{ALFA and NaCo observations of \object{HH\,Leo} A \& BC}
\label{obsec}

\begin{figure}
\centering
\resizebox{0.99\hsize}{!}{\includegraphics[]{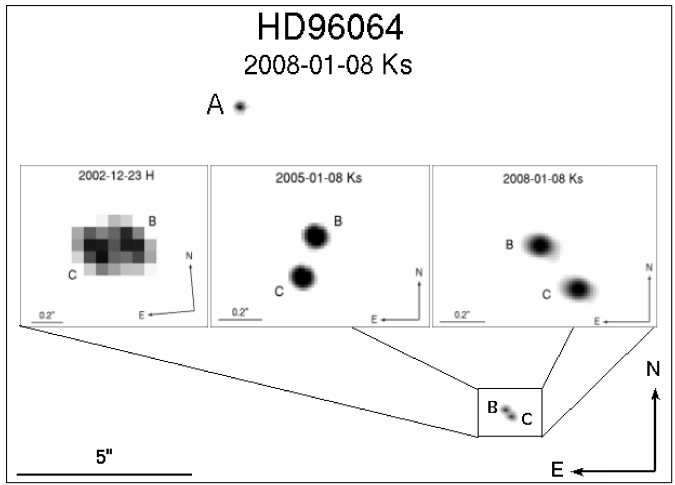}
}
\caption{The \object{HH\,Leo} (= \object{HD\,96064}) system from 2002 (Calar Alto $\Omega$-Cass with ALFA, inset left), through 2005 (inset middle), to 2008 (inset right and main image, both VLT/NaCo). Orbital motion of the BC component can clearly be seen. For further explanations see text. }
\label{hhleo}
\end{figure}

We observed \object{HH\,Leo} (\object{HD\,96064}) in 2002 December 23 with $\Omega$-Cass and the adaptive optics system ALFA on Calar Alto. We used the shortest possible DIT of 0.842\,s and 49 integrations were averaged to one frame per jitter position. We took images at 20 jitter positions, i.e., total integration times of 13.8\,min. The data were reduced in a standard way and calibrated using the calibration provided in \citet{2004A&A...417.1031M}. We resolved  the binary companion $\sim11''$ to the southwest of \object{HH\,Leo} (Fig.~\ref{hhleo}). Both, B and C, are listed in SIMBAD as objects \object{LTT\,4076} and BD-033040C, respectively. It is an equal mass binary orbiting \object{HH\,Leo}\,A. However, the objects are typically not resolved in most photometric catalogs. Since the information found for this system is often part of a compilation of huge data sets, it is unlikely that the multiplicity of tight systems was treated correctly. In that sense, the magnitudes and colors given for both objects in the literature have to be used with care. 

\begin{table}
\caption{Separation and position angle of \object{HH\,Leo} B \& C from 1989 to 2008.  }
\label{hhleoobslog}
{\scriptsize
\begin{center}
\begin{tabular}{ll@{$\ \pm\ $}lr@{$\ \pm\ $}ll}
\hline\hline
Epoch	&	\multicolumn{2}{c}{separation}	&	\multicolumn{2}{c}{position angle}	&	Reference\\
         	&	\multicolumn{2}{c}{['']}	&	\multicolumn{2}{c}{[$^{\circ}$] }     	&	                 \\
\hline
1989 Nov 10&  0.342    &	0.0027&	  257.2&	0.4 * &       \citep{2000AJ....119.3084H}\\
1990 Apr 03&  0.313    &	0.009  &	266.5&	0.6  &       \citep{1993AJ....106..352H}\\
1991 Apr 20&  0.264    &	0.003  &	272.3&	0.2  &       \citep{1993AJ....106..352H}\\
1993 Jan 02&  0.175    &	0.0014&	305.4&	0.4  *&       \citep{2000AJ....119.3084H}\\
1997 Mar 06&  0.280    &	0.0022&	54.3&	0.4  *&       \citep{2000AJ....119.3084H}\\
1999 Oct 19&  0.311    &	0.0028&	83.7&	0.91 *&       \citep{2002AJ....123.3442H}\\
2002 Dec 23&	0.247  &	0.010  &	112.7&	0.10&this work\\
2005 Jan 08&	0.177&	0.0014&	161.9&	0.64&this work\\
2008 Jan 08&	0.254&	0.004&	220.9&	0.13&this work\\
\hline
\end{tabular}\\
\end{center}
\tablefoot{
*: original values corrected by $\pm 180^{\circ}$ for consistency\\
Data are obtained via speckle interferometry (until 1999) and direct imaging. }
}
\end{table}

Within our direct imaging campaign of sub-stellar companions, we observed \object{HH\,Leo} again with VLT/NaCo \citep{2003SPIE.4841..944L,2003SPIE.4839..140R} in 2005 and 2008. The observed target was also used for the wavefront sensing. The observations were performed in the jitter mode with the shortest possible detector integration time (DIT) of 0.347\,s, and 200 such integrations were always averaged to one frame per jitter position. In the 1st epoch 27 frames, and in the 2nd epoch 15 frames are taken at different jitter positions, which results in total integration times of 31.2\,min and 17.4\,min, respectively. We calibrated the NaCo data with the well-known binaries AB\,Dor and HIP\,73357, respectively. Data reduction was  done with {\it ESO Eclipse}. The resulting separation and position angle of the equal-mass binary system is shown in Table~\ref{hhleoobslog} and orbital motion is visible in Fig.~\ref{hhleo}. Realizing that we could determine the orbit more precisely, we searched the literature and found six speckle interferometry measurements from 1989 to 1999, see \citet{1993AJ....106..352H,2000AJ....119.3084H}, \citet{2002AJ....123.3442H}, and Table~\ref{hhleoobslog}.
Together with our observations, this is almost a full orbit which is suspected to be $\sim 23\,$years. Because of the similar brightness of both objects, the B and C components were not commonly defined in the literature. We therefore adopt a correction to the position angle of $180^\circ$ where necessary.

\begin{figure*}
\centering
\vspace*{1cm}
\begin{minipage}[!]{10cm}{\subfigure{\includegraphics[height=11cm]{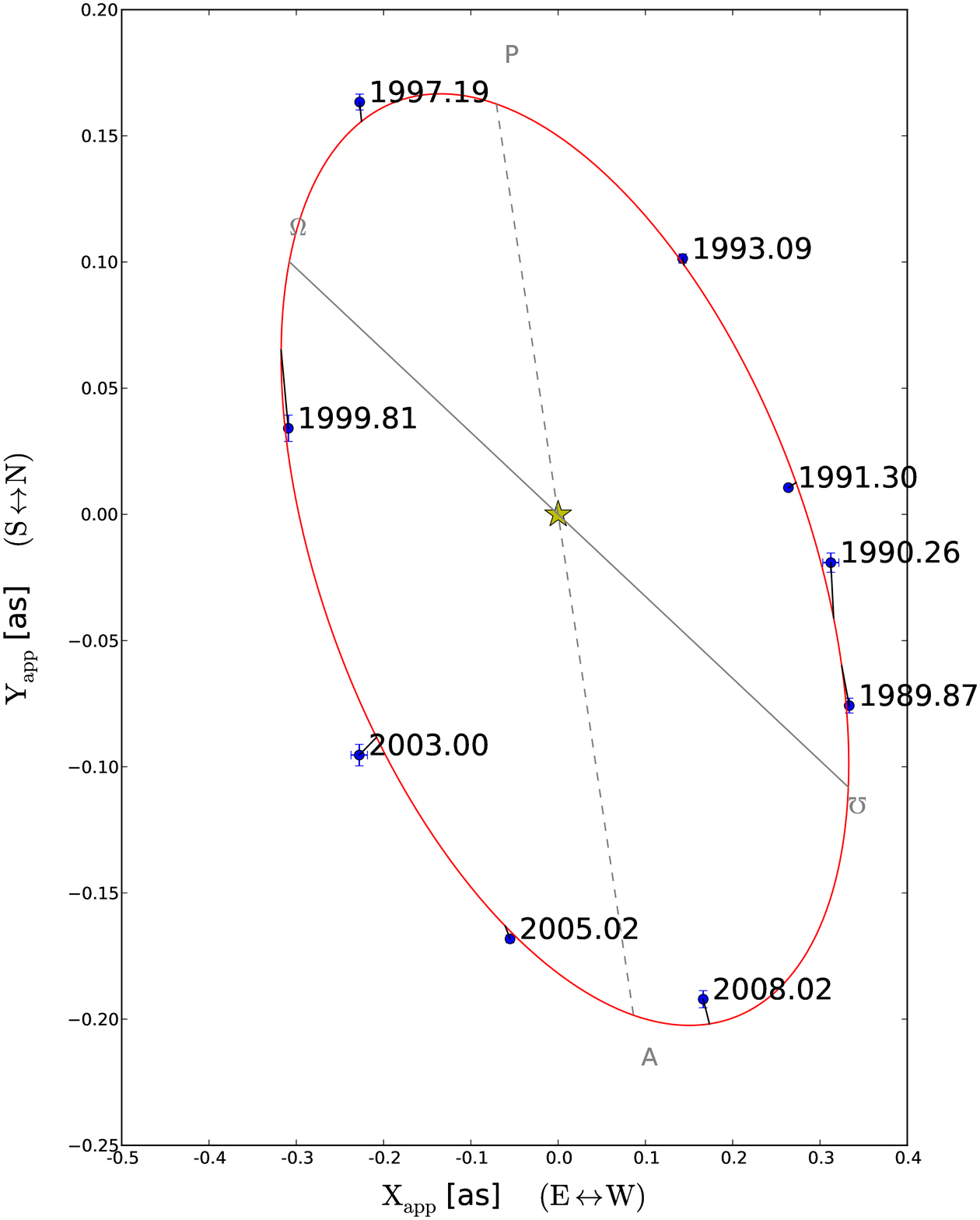}}}
\end{minipage}
\begin{minipage}[!]{8cm}
\subfigure{\includegraphics[clip,width=8cm,height=5.5cm]{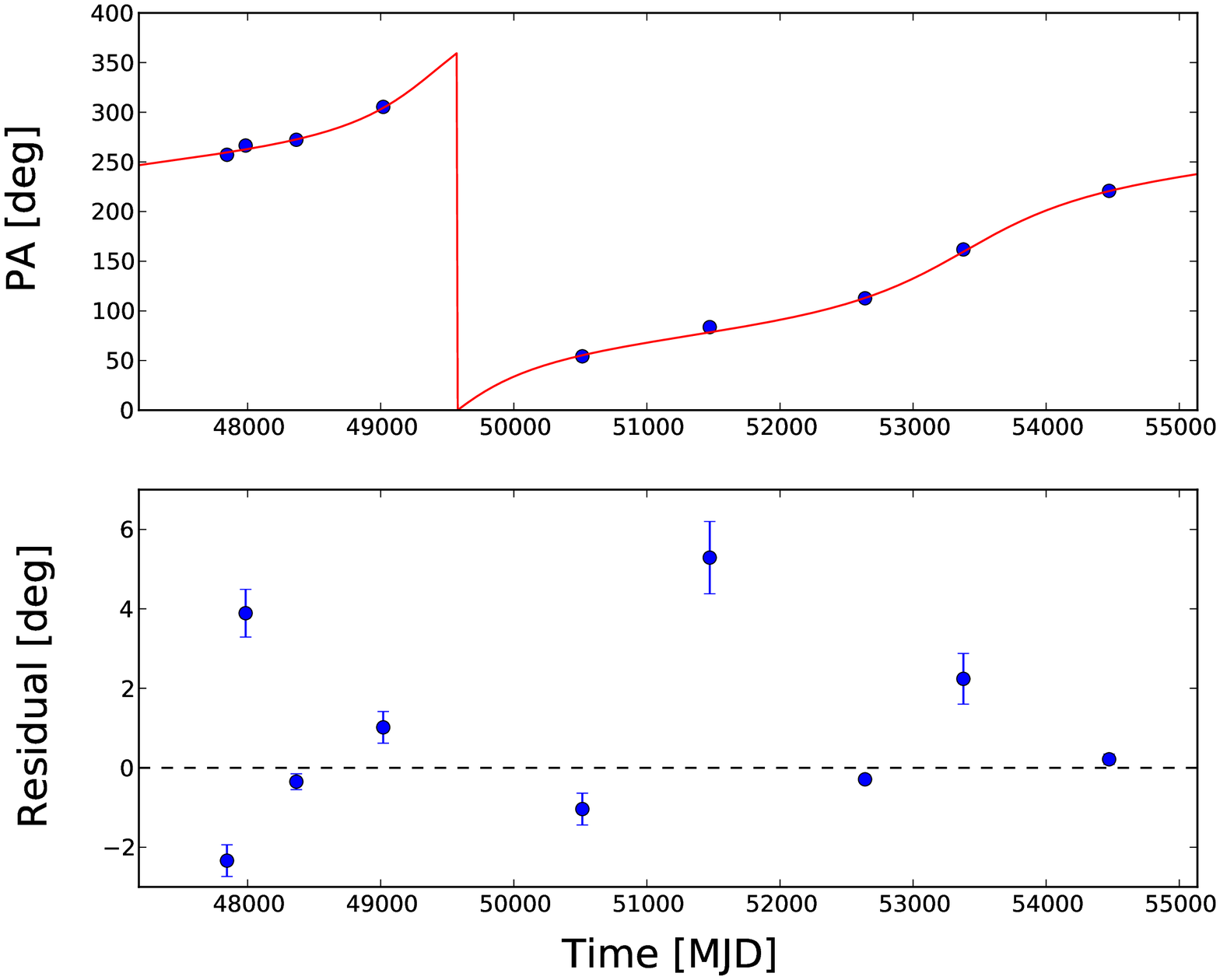}}
\subfigure{\includegraphics[clip,width=8cm,height=5.5cm]{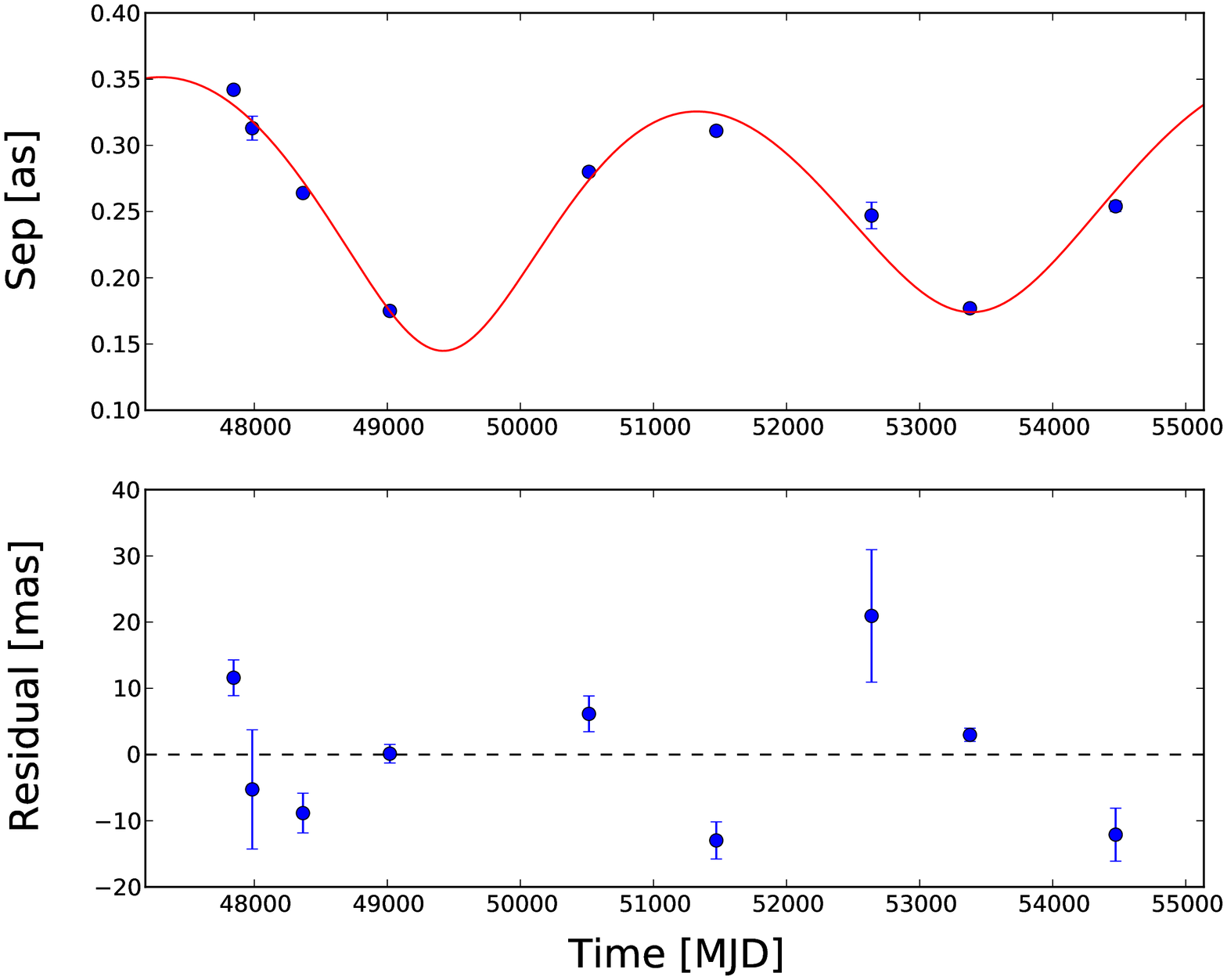}}
\end{minipage}
\caption{Orbital solution for the HH\,Leo B \& C binary system obtained by a $\mathrm{\chi^2}$ minimization. The separation (in arcsec) and position angle measurements as well as their residuals to the binary model are displayed on the right side.}
\label{hhleoorb}
\end{figure*}

Using this dataset of nine observations (18 constraints: nine separations and nine position angles), we fit the orbit of HH\,Leo B\,\&\,C (see Fig.~\ref{hhleoorb}) by using an error weighted $\mathrm{\chi^2}$ minimization using the {\it SciPy} package \citep{python_scipy}. The $\mathrm{\chi^2}$ is calculated as
\begin{equation}
 \mathrm{\chi^2 = \sum_i \frac{(O_i-C_i)^2}{\sigma_i^2},}
\end{equation}
{  where} $\mathrm{O_i}$ are the measured values,  $\mathrm{C_i}$ the  model values, and $\mathrm{\sigma_i}$  the measurement errors. 

The distance of the system was fixed to $\mathrm{26.27\,pc}$, taken from \citet{2007A&A...474..653V}. Although HH\,Leo is a hierarchical triple star, the influence of the A component can be neglected and HH\,Leo B\,\&\,C can be treated as a binary system. The achieved orbital parameters of the HH\,Leo B\,\&\,C system are listed in Table \ref{oftbl} together with their $\mathrm{1\sigma}\,$- and $\mathrm{3\sigma}\,$-uncertainties which are determined by varying the fit parameters \cite[see][for further details]{2002nrc..book.....P}.

{With seven orbital parameters to fit, {  there are ten degrees of freedom} in total. The achieved reduced $\mathrm{\chi^2}$ of the best orbital fit is $22.63$ which is quite large and also {  obvious in view of} the large residuals of the separation and position angle measurements (see Fig.~\ref{hhleoorb}). Assuming that no systematic effects affected the measurements, there are two possible explanations for  residuals { that large}.

First, the errors of the speckle interferometry measurements could be underestimated because \citet{1993AJ....106..352H,2000AJ....119.3084H} published no more than the typical errors for each telescope of the observation program \citep[see][Table 3]{2000AJ....119.3084H}, but not the measurement errors for a specific target system itself. \citet{2002AJ....123.3442H} also published average errors for their observation campaign. The second possibility is the presence of an unseen astrometric companion in the HH\,Leo\,B\,\&\,C system. However, attempts to fit a model of a binary plus an astrometric companion around one of the stellar components did not result in smaller residuals. In fact, the $\mathrm{\chi^2_{red}}$ increases, especially because of the reduced degree of freedom. Hence, the existence of an unseen astrometric companion could neither be confirmed nor ruled out with the present data of nine observations. Additional observations are needed. The values and errors in Table 4 are based on the assumption {that the real system can be described by a binary model}. 
}

\begin{table}
\caption{\label{oftbl} Parameters of the binary system HD\,96064 B\,\&\,C calculated by a $\chi^2$ minimization.}
\renewcommand{\arraystretch}{1.5}
\begin{center}
\begin{tabular}{lcc}
\hline\hline
Parameter & $\mathrm{1\sigma - uncertainty}$ & $\mathrm{3\sigma - uncertainty}$ \\ \hline 
$\varepsilon$	&	$0.098^{\,+0.003}_{\,-0.003}$	&	$0.098^{\,+0.008}_{\,-0.008}$\\ 
$\omega$ [$^\circ$]	&	$292.603^{\,+0.240}_{\,-0.160}$	&	$292.603^{\,+0.641}_{\,-0.561}$\\ 
$M_\mathrm{tot}$ [$M_\odot$]	&	$1.353^{\,+0.041}_{\,-0.041}$	&	$1.353^{\,+0.119}_{\,-0.126}$\\ 
$P$ [d]	&	$8410.460^{\,+14.615}_{\,-14.615}$	&	$8410.460^{\,+43.844}_{\,-44.044}$\\ 
$T$ [MJD]	&	$49843.343^{\,+5.506}_{\,-5.005}$	&	$49843.343^{\,+16.016}_{\,-15.516}$\\ 
$d$ [pc]	&	$26.194^{\,+0.260}_{\,-0.264}$	&	$26.194^{\,+0.765}_{\,-0.809}$\\ 
$\Omega$ [$^\circ$]	&	$71.872^{\,+0.160}_{\,-0.240}$	&	$71.872^{\,+0.641}_{\,-0.721}$\\ 
$i$ [$^\circ$]	&	$62.022^{\,+0.200}_{\,-0.240}$	&	$62.022^{\,+0.681}_{\,-0.721}$\\ \hline
\end{tabular}
\end{center}
\tablefoot{The listed uncertainties are determined according to \citet{2002nrc..book.....P}.}
\end{table}

In the 2MASS catalog, $K_\mathrm{s}$ magnitudes of $5.801\pm0.021$\,mag and $6.416\pm0.016$\,mag are given for the A and the unresolved BC component, respectively. We measured the instrumental magnitude of the A component in our NaCo images to determine the zero point with an aperture radius of 27\,pixels. Then, we measured the B and C component individually with a smaller aperture of 18\,pixels. In addition, both B and C were measured within a single, larger aperture of 55 pixels. 

\begin{figure}
\centering
\resizebox{1.0\hsize}{!}{\includegraphics{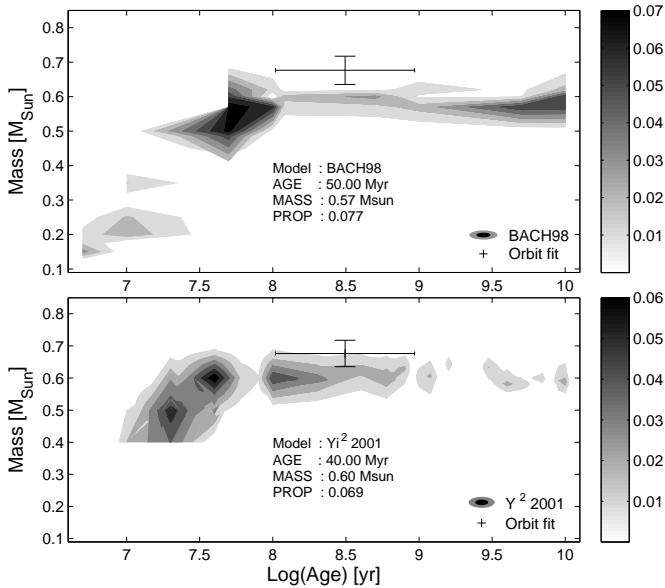}}
\caption{Surface plot of the results of a Monte-Carlo simulation of a comparison of the photometric measurements ($M_{K_\mathrm{s}}$ from NaCo and $J-H$ from 2MASS) and evolutionary models by \citet{1998A&A...337..403B} and \citet{2001ApJS..136..417Y}. Large probabilities are shown in dark colors and low probabilities are shown in light colors. The color bar on the right show the encoded probability. For the highest probability, the age and the mass are given in the corresponding subfigures. The datapoint shows the dynamical mass of HH\,Leo\, B$ + $C, divided by 2, and the { gyrochronological} age of HH\,Leo for comparison.}
\label{hhleomass}
\end{figure}

The apparent magnitudes of B, C, and the combined magnitude of BC are derived by aperture photometry relative to the well-known A component. 
 Using the parallax of A, we get the absolute magnitude $M_{K_\mathrm{s}}$. This measurement indicates that both objects are nearly similar in brightness and may also have a similar spectral type. Although this assumption cannot be proven directly, we can now compare our dynamically determined mass to evolutionary models. As in Sect. \ref{photo}, we use the models by \citet{1998A&A...337..403B} and \citet{2001ApJS..136..417Y} for this comparison. Additional photometric information is taken from the 2MASS catalog. We utilized our measurement of the absolute $K_\mathrm{s}$ magnitude and the 2MASS $J-K$ color and ran a Monte Carlo simulation of 1 million runs for each model to find the most probable mass and age. The result is shown in Fig.~\ref{hhleomass}. For comparison, the total dynamical mass divided by two and the { gyrochronological} age (see Sect. \ref{gyro}) is shown as a data point. Figure~\ref{hhleomass} shows that both models estimate the mass of the object to be approximately $0.6\,M_{\odot}$ which is below but approximately in agreement with half the dynamical total mass of the system.  The maximum probability isochronic age however is 40-50\,Myr in disagreement with the { gyrochronological} age of $\approx 300\,$Myr. However, the age is not very well constrained by the models.

The photometric measurements are given along with the model based spectral types and masses \citep{1998A&A...337..403B} in Table \ref{magtab}. 
It turns out that the total mass of B+C is higher than the mass of component A.

\begin{table}
\caption{\label{magtab} $K_\mathrm{s}$ band photometry of HH\,Leo A \& BC as extracted out of our VLT/NaCo images using relative aperture photometry. }
\begin{center}
\begin{tabular}{ccclc}
\hline\hline
comp. & $K_\mathrm{s}$ & $M_{K_\mathrm{s}}$ & SpT & Mass \\
	   & mag & mag &    & $\Msun$\\
\hline  
A	& $5.801\pm0.021$ & $3.84\pm0.193$	&G5&0.92\\
BC	& $6.239\pm0.021$	&                          	&&$\approx 1.2$\\
B	& $7.157\pm0.022$	& $5.20\pm0.194$	&K7 - M0& $\approx0.6$\\
C	& $7.197\pm0.022$	& $5.24\pm0.194$	&K7 - M0& $\approx0.6$\\
\hline
\end{tabular}
\end{center}
\tablefoot{The 2MASS catalog serves as a calibration source. The masses are photometric  \citep{1998A&A...337..403B}.}
\end{table}

\subsection{Summary of all Her-Lyr candidates}
{ After collecting all candidates from the literature and performing the multiplicity study, as many as 48 visually resolved Her-Lyr candidates are taken into account in this analysis. In addition,  there are three brown dwarf companions, one for \object{HN\,Peg} \citep{2007ApJ...654..570L} and a brown dwarf binary companion to \object{MN\,UMa} \citep{2003AJ....126.1526B}. These brown dwarfs are listed in Tables \ref{hltab} and \ref{hltab2} but are not counted separately. Furthermore, three membership candidates (\object{BC\,Ari}, \object{HD\,54371}, and \object{PX\,Vir}) are known spectroscopic binaries. The complete census of Her-Lyr candidates amounts to 54 individual objects, distributed over 35 stellar systems. This work is aimed at the 48 stellar objects and does not discuss the spectroscopic binaries in detail. For more information on individual objects, please refer to Sect. \ref{indstar} and Tables \ref{hltab} and \ref{hltab2}.}

\section{Photometry}

\label{photo}
\begin{figure*}
\centering
\subfigure[$Y^2$ isochrones for solar type stars]{\includegraphics[width=9cm]{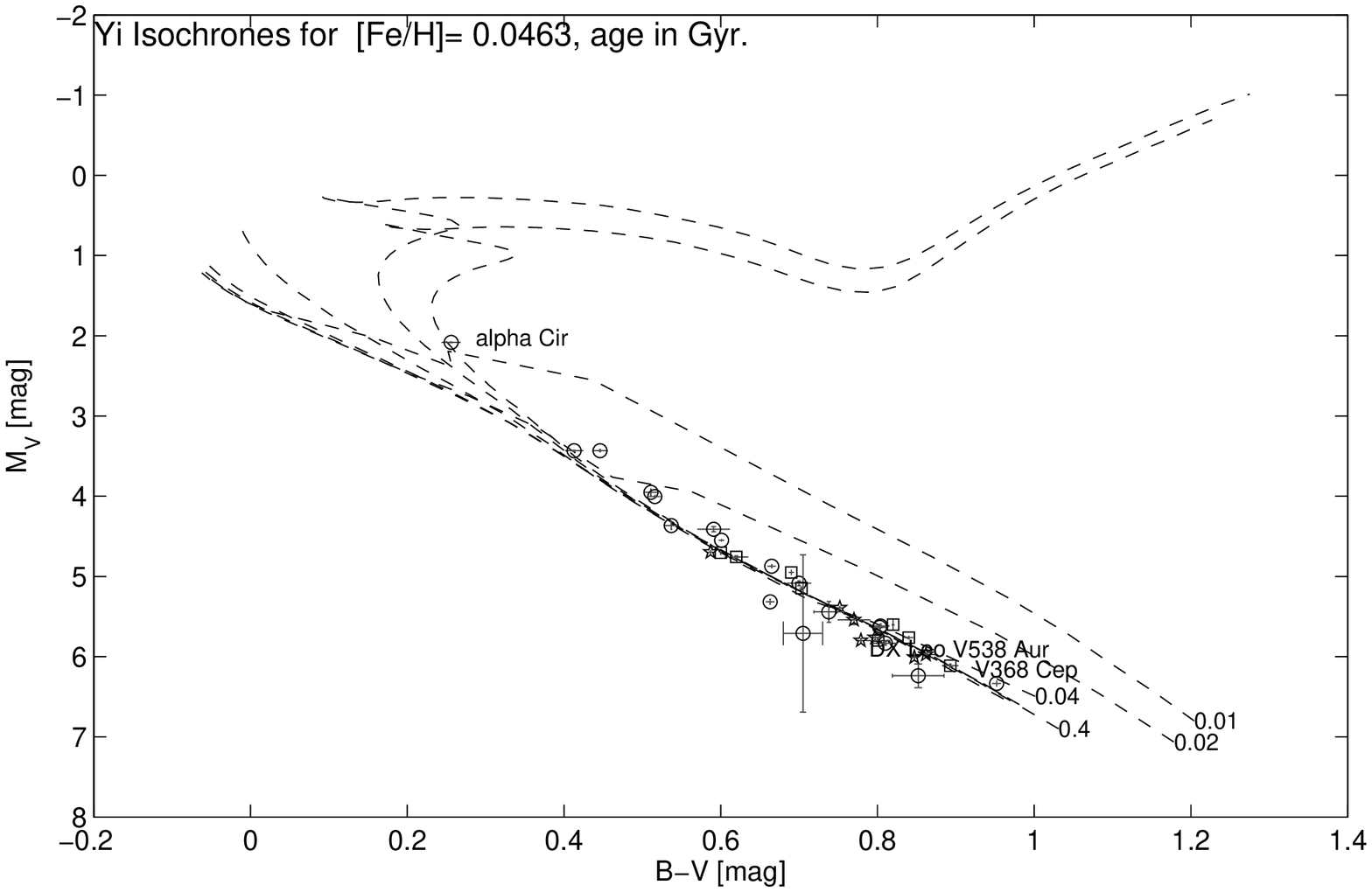}}
\subfigure[Baraffe isochrones for brown dwarfs]{\includegraphics[width=9cm]{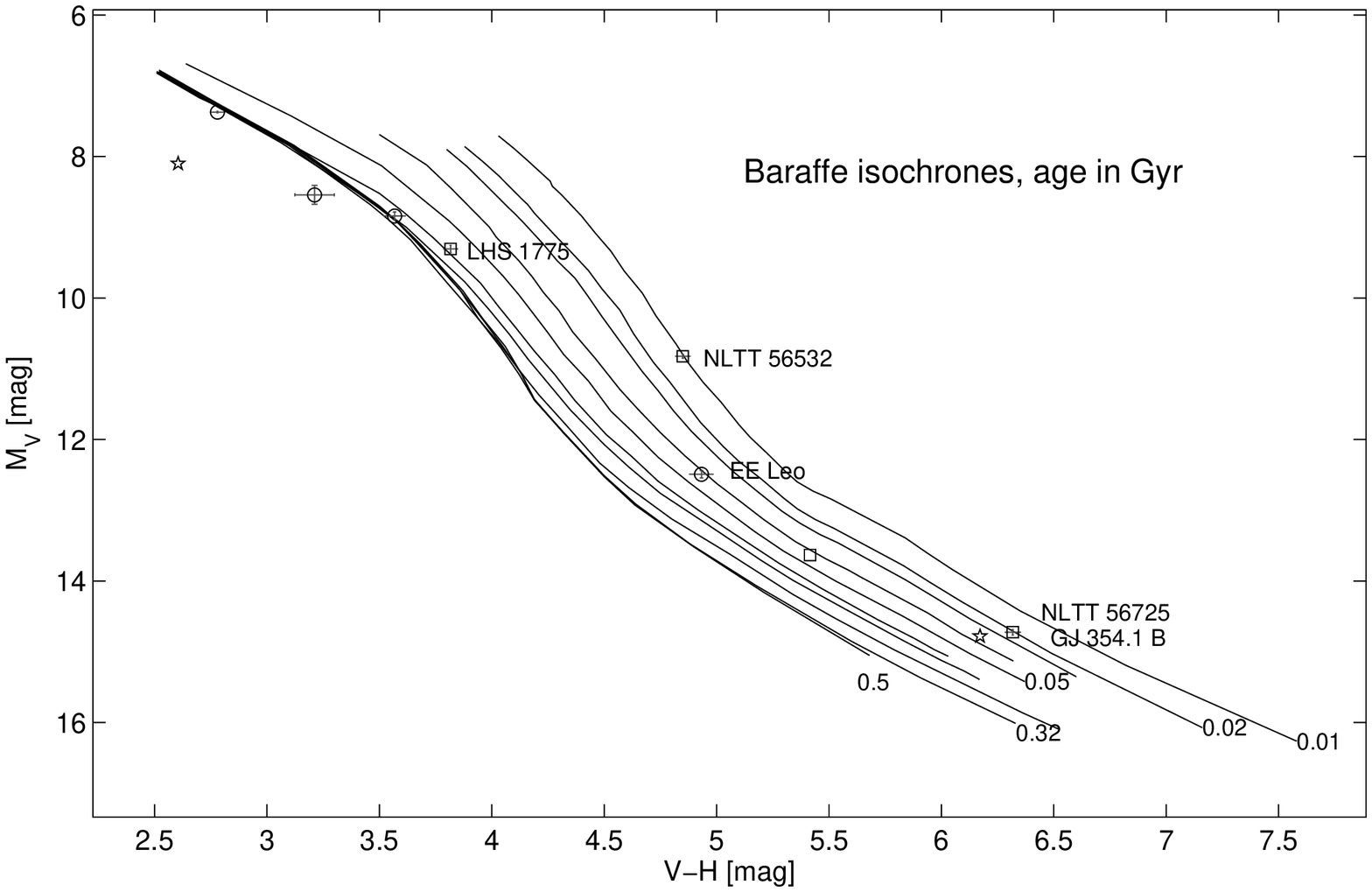}}
\caption{\label{photfig}(a) The FGK-type Her-Lyr candidates compared with model isochrones by \citet{2001ApJS..136..417Y}, the $Y^2$ isochrones for solar mixture ($Z=0.02$, $Y=0.27$, $OS=0.20$, $l/H_p=1.743201$ $[\mathrm{Fe}/\mathrm{H}]= 0.046320$, and  $[\alpha/\mathrm{Fe}]= 0.00$). {$M_V$ is displayed vs. $B-V$ which is a well-calibrated temperature indicator for FGK-type stars.}
(b) The M-type Her-Lyr candidates compared with isochrones by \citet{1998A&A...337..403B}, optimized for late-type stars and brown dwarfs. The $V-H$ color index is used for the late-type stars. 
\newline
 The photometry is taken from the HIPPARCOS catalog \citep{1997A&A...323L..49P}, 2MASS \citep{2003tmc..book.....C}, and other catalogs if HIPPARCOS data were not available \citep{2000A&AS..143...23O}. The full list of references is given below Table~\ref{hltab}. According to Table~\ref{memtab}, stars denote {probable} Her-Lyr members, circles denote uncertain {members}, and squares denote non-members. In the case of stars shown without error bars, uncertainties were not given in the corresponding catalog. For conversion from apparent to absolute magnitude, we used the revised parallaxes from \citet{2007A&A...474..653V}. For faint companions, we adopted the parallax of the primary. 
 }
\end{figure*}

\begin{figure}
\centering
\resizebox{1.0\hsize}{!}{\includegraphics{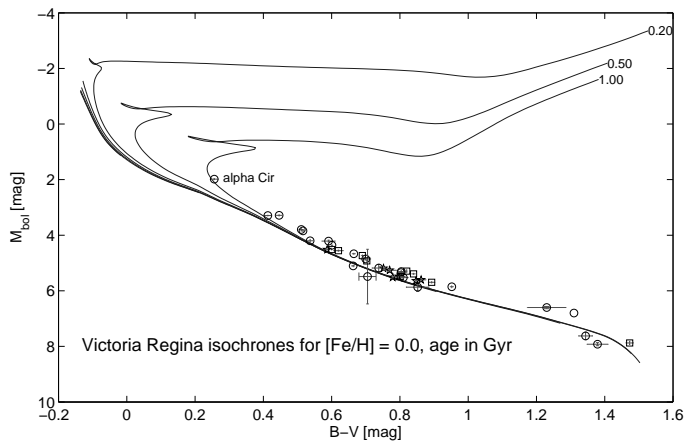}}
\caption{{ Victoria Regina isochrones based on evolutionary models reported in \citet{2006ApJS..162..375V}. The models include $\alpha$ enhancement and core overshooting. Stellar mass ranges from $0.4$ to $4.0\, M_{\odot}$ and stellar age from $0.2$ to $1.0$\,Gyr. Symbols denote Her-Lyr candidates as described in the caption to Fig.~\ref{photfig}.}}
\label{vafig}
\end{figure}

Since most of the Her-Lyr members and even the late type members are sufficiently bright {$B-V$ and $V-H$ data are available from the} HIPPARCOS \citep{1997A&A...323L..49P} and the 2MASS catalog \citep{2003tmc..book.....C}. {These data can be compared} with {evolutionary} models. The late-type companions may contribute important information to a photometric age estimation, based on evolutionary models since late-type stars arrive later on the main sequence than early-type stars. { We have chosen the models by \citet{1998A&A...337..403B}, \citet{2001ApJS..136..417Y} (Fig.~\ref{photfig}), and \citet{2006ApJS..162..375V} (Fig.~\ref{vafig})}. { Giving a good representation of the main sequence in the regime of solar-type stars, the models disagree for late-type stars.} 

Examining Fig.~\ref{photfig} (a) gives very limited insight into the age of the association. Most membership candidates are located on the zero age main sequence (ZAMS), the late end being at spectral type K7 to M0. So from Fig.~\ref{photfig} (a) alone, it can be concluded that the Her-Lyr association is $\gtrsim40$\,Myr old.

Figure~\ref{photfig} (b) on the other hand shows the M dwarfs of the sample. Most of them are companions of brighter Her-Lyr stars. 
All  stars { except two} down to M1 are located on or near the ZAMS of the Baraffe isochrones. The five mid- to late-M stars of the sample are only visible in Fig.~\ref{photfig} (b). Only the \citet{1998A&A...337..403B} isochrones are extended to their mass range, implying a young age for all five.

{  Figure~\ref{vafig} shows the model grid of \citet{2006ApJS..162..375V} intended to fit the ZAMS and the evolution of stars for a wide range of masses and ages. The isochrones of these models indeed coincide with most of the Her-Lyr candidates. For this figure, we calculated the bolometric magnitude using the corrections given in \citet{1995ApJS..101..117K}. The bolometric correction is given in Table \ref{hltab2}. Figure~\ref{vafig} indicates that most of the Her-Lyr candidates are coeval down to $0.4\,M_{\odot}$. They are located above the isochrone of $0.2\,$Gyr implying a young age. As the isochrones of \citet{2006ApJS..162..375V} do not include pre-main sequence tracks they do not provide further information about the age.}

Lessons to take away from Fig.~\ref{photfig} and Fig.~\ref{vafig} are:
\begin{itemize}
\item
It is not clear if any of the models predict correct colors for M stars. If they are correct, Fig.~\ref{photfig} (b) shows significant age spread among the M-type candidates.
\item
\object{EE\,Leo} is a poorly known isolated M dwarf. No spectroscopic measurements were found for this star to confirm or disprove that this is indeed a young object. \object{Gl 354.1\,B} is the companion of \object{DX\,Leo} which was thought to be younger than the average of Her-Lyr stars because of its strong activity \citep{2006ApJ...643.1160L}. The stars { \object{NLTT\,56532} and \object{NLTT\,56725}} are companions to \object{V368\,Cep}. There are many indicators that this star (and its companions) is very young (12-50\,Myr) and, therefore, a doubtful member of the Her-Lyr association. It cannot be excluded that the selection of M-type objects for this study was composed almost entirely of non-members.
 
\item 
As presented in Sect.~\ref{multi}, \object{V538\,Aur} and \object{LHS\,1775} form a binary. Compared with the  \citet{2001ApJS..136..417Y} isochrones, \object{LHS\,1775} is located exactly on the $40\,$Myr isochrone. However, as seen from Table~\ref{hltab}, the system seems to be  older than the Her-Lyr association implying  an inconsistency between evolutionary models and other age indicators.
\item
The A7 VpSrCrEu star \object{$\alpha$\,Cir}, visible in Fig.~\ref{photfig} (a), is either very young ($<10\,$Myr) or very old ($\sim 1$\,Gyr) as it { is located} close to the intersection of both isochrones. It is the brightest rapidly oscillating Ap star and was studied in detail by \citet{2009MNRAS.396.1189B}. In Fig.~\ref{vafig}, $\alpha$\,Cir is located exactly on the $1$\,Gyr isochrone suggesting that this object might be older than Her-Lyr but not a member. { The absence of isochrones younger than $0.2\,$Gyr implies that a very young age for $\alpha$\,Cir cannot be ruled out.}
If it is not a member of the Her-Lyr association, its companion \object{LTT\,5826} is also not a member.
\end{itemize}

 In fact, evolutionary models cannot be used to rule out non-members or to estimate the association's age. This is partly caused by measurement inaccuracies and the density of the isochrones near the ZAMS, and partly by the fact that a valid representation of equally aged stars from late M-dwarfs to A stars is still difficult to determine from stellar evolution theory.

\section{Youth indicators}
\label{youth}
{ The most obvious} evidence of a young Her-Lyr association is the concentration of late-type fast rotators in kinematic space. Given the volume-completeness by \citet{2004AN....325....3F} and the absence of any pre-selection, this is an unbiased view. Therefore, this concentration in kinematic space provides a solid starting point to critically review Her-Lyr membership. In this work, all kinematic candidates \citep{1998PASP..110.1259G,2004AN....325....3F,2006ApJ...643.1160L,2008MNRAS.384..173F} are revisited without any pre-selection based on lithium, activity, or photometry. The strategy is to study the distribution of these properties of all kinematic candidates. A population of young stars will stand out clearly in this distribution. In particular, this is expected for the rotational velocities since these gave rise to the {notion} that the concentration in kinematic space is a group of relatively young stars \citep{2004AN....325....3F}.

\subsection{The lithium equivalent width}
\label{lithium}
\begin{figure}
\centering
\resizebox{1.0\hsize}{!}{\includegraphics{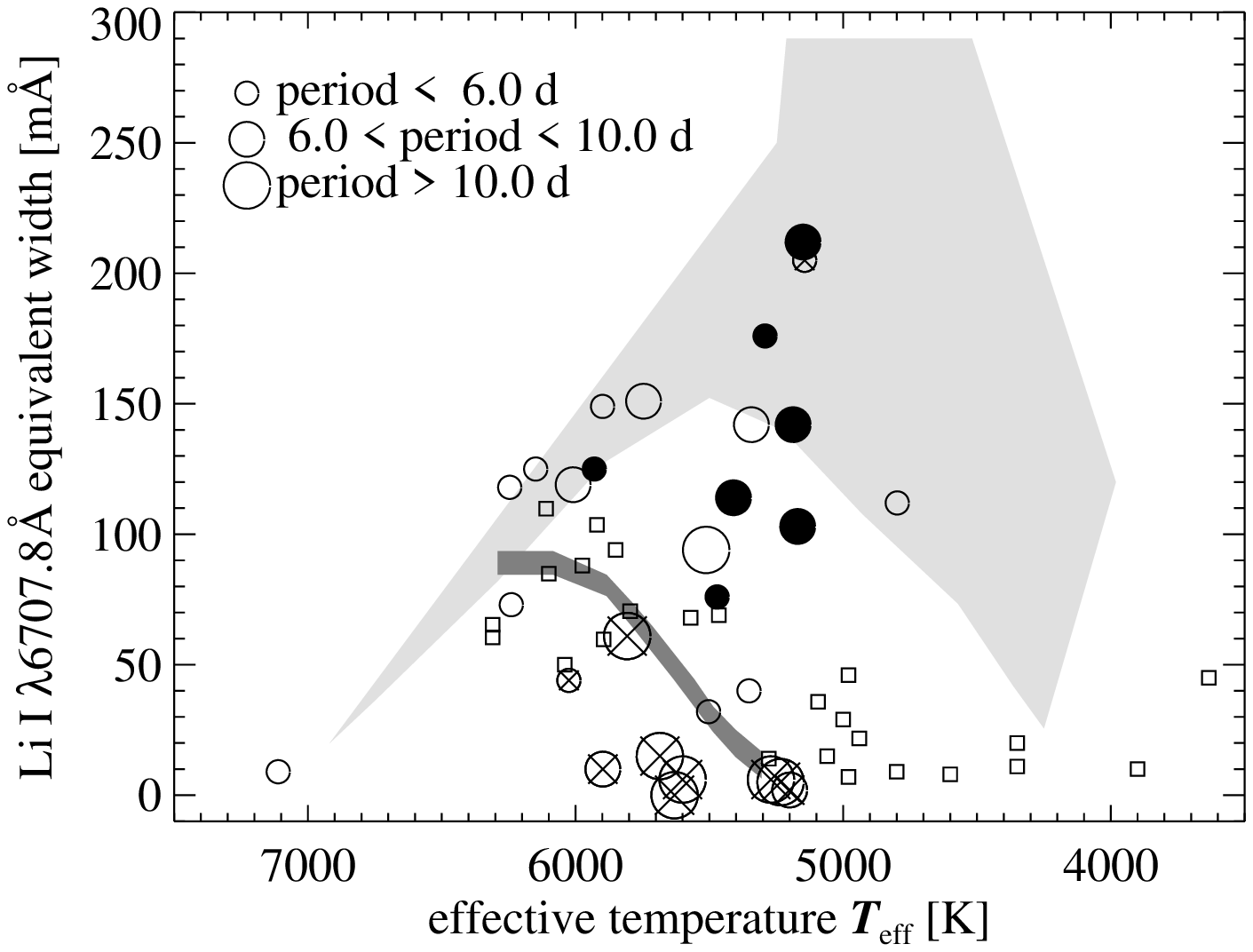}}
\caption{Lithium equivalent widths of Her-Lyr candidates compiled in this work (circles) compared to the areas populated by the Pleiades (gray, \citealp{1997A&A...325..647N,1993AJ....106.1059S}) and Hyades (dark gray, \citealp{1990AJ.....99..595S}). UMa group members \citep{2009A&A...508..677A} are indicated by squares. Effective temperatures and lithium equivalent widths of Her-Lyr candidates are taken from Table~\ref{hltab}. Error bars  have been omitted for clarity . { The symbol sizes of Her-Lyr stars scale with rotational period as is indicated in the figure. Symbols denote final membership according to Table \ref{memtab}.
Filled circles identify members (+) while doubtful (o) or possible members (?) are highlighted by open circles. Crossed circles denote
rejected (-) candidates.}}
\label{lifig}
\end{figure}

Figure~\ref{lifig} compares the Li equivalent widths of all candidates to the distribution of the Pleiades, UMa, and the Hyades. The upper envelope of the Pleiades is taken from \citet{1997A&A...325..647N} while the lower envelope is obtained from \cite{1993AJ....106.1059S}.  The UMa data are adopted from \citet{2009A&A...508..677A} { providing a homogeneous analysis}. The Hyades region originates from \citet{1990AJ.....99..595S}. The colors given by \citet{1993AJ....106.1059S} and \citet{1990AJ.....99..595S} are converted to effective temperatures using the scales of \citet{1979PASP...91..589B} and \citet{1991AJ....101..662B}. The comparison is limited to effective temperatures above $\approx 5000$\,K {since} lithium measurements are not available for the cooler Her-Lyr candidates. The data of the Her-Lyr candidates { form two distinct groups.} One group displays lithium absorption at the level of the Pleiades and the UMa group while the other group does not show any significant lithium absorption.

{  Excluding} the lithium-depleted stars, the Her-Lyr candidates show at least as much lithium as the UMa group but not more than the Pleiades, indicating that they are young stars with an age not very different from the Pleiades and the UMa group. The age of Her-Lyr without the Li-depleted candidates is therefore similar to UMa, or between Pleiades and UMa and younger than the Hyades.

\subsection{Rotation}
\label{gyro}

It is well known that stellar rotation slows down as a star becomes older. However, this effect varies with stellar mass and radius as well as with starting conditions, accretion history, and disk life time. 
Therefore, it is possible to roughly estimate the age of a single main sequence star from its rotational period by taking into account its spectral type or its $B-V$ color.

An empirical relation was found by \citet{2007ApJ...669.1167B} and was improved by \citet{2008ApJ...687.1264M} and \citet{2009IAUS..258..375M} {to be}
\begin{equation}
\log t= \frac1n \left (\log P - \log a -b\,\log(B-V-c) \right )\label{gyroeq}
\end{equation}
{with $t$ being} the age in {  millions of years} and $P$ the rotational period in days.

\begin{table}
\caption{Parameters for Eq. \ref{gyroeq} as estimated by various authors. }
\label{gyrotab}
\begin{tabular}{cccc}
\hline\hline
&\citet{2007ApJ...669.1167B}&\citet{2009IAUS..258..345B}&\citet{2009IAUS..258..375M}\\
\hline
$n$		&$0.5189\pm0.007$	&$0.519\pm0.007$		&$0.566\pm0.008$	\\
$a$		&$0.7725\pm0.011$	&$0.770\pm0.0014$	&$0.407\pm0.021$	\\
$b$		&$0.601\pm0.024$	&$0.553\pm0.052$		&$0.325\pm0.024$	\\
$c$		&$0.4$			&$0.472\pm0.027$		&$0.495\pm0,010$	\\
$(B-V)$	&$>c$			&$>c$				&$0.5\ldots0.9$	\\
\hline
\end{tabular}
\tablefoot{The last {row gives the applicable} range of $(B-V)$.}
\end{table}

The sets of parameters for Eq. \ref{gyroeq} are summarized in Table~\ref{gyrotab}. In the analysis described in the following, we use either the more recent values of \citet{2009IAUS..258..375M}, if possible, { or} the values of \citet{2009IAUS..258..345B} if $B-V$ is outside the {applicable} range but still larger than $c$. If $(B-V)< 0.472$ but still larger than 0.4, we use \citet{2007ApJ...669.1167B}.  Uncertainties in all cases are calculated following  \citet{2009IAUS..258..345B}. It should also be mentioned that this method is not applicable to mid- and late-M dwarfs as {these form a fully convective envelope leading to} a different rotational behavior.

In principle, the rotational period can be derived in two ways. The direct way is to observe a photometric light curve. If periodic brightness variations caused by star spots {are observed}, the photometric periodicity is a direct measure of the rotational period. A possible complication might be the observation of multiples of the rotational period if several spots are located at unfavorable stellar longitudes.

On the other hand, the broadening of sufficiently resolved {  spectral} lines gives a measure of the rotational velocity, but multiplied by the sine of the inclination. The rotational period then is $P/\sin i = 2 \pi R / v\sin i$ with the stellar radius $R$. This introduces an additional uncertainty. We use the photometric period, if available, or the spectroscopic period. In cases where both are available, we { confirmed} that the differences are generally small.

\begin{figure}
\centering
\resizebox{1.0\hsize}{!}{\includegraphics{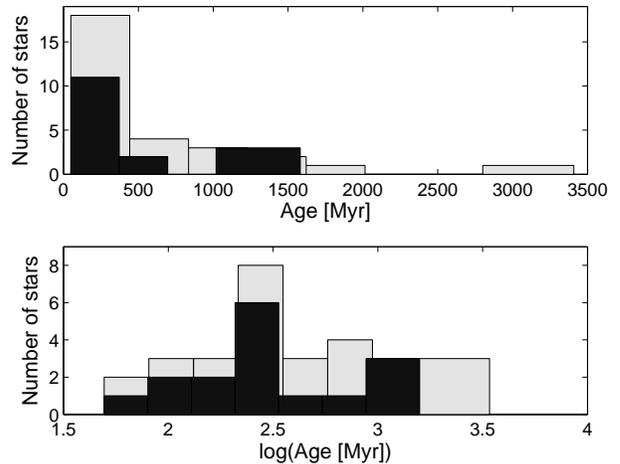}}
\caption{ Age distribution of Her-Lyr candidates derived from the rotational period (black).  In the gray data, rotational periods assessed from $v\sin i$ measurements are included. From these measurements, a lower limit of the rotational speed is derived, leading to an upper limit of the { gyrochronological} age. In the top panel, the age is displayed on a linear scale, emphasizing the existence of a young concentration of 11 stars (18 including upper limits). In the lower panel, the age is plotted logarithmically to illustrate that the concentration of young stars consolidates with six (eight) stars with similar age. 
The binning of the histograms employs the Freedman-Diaconis rule \citep{springerlink}.}
\label{hlfig}
\end{figure}

{Photometric rotational periods are available for 16 Her-Lyr candidates with spectral types F, G, and K (see Table \ref{hltab}).} 
Figure~\ref{hlfig} shows the distribution of the ages derived for these 16 stars in linear and logarithmic scale. It can be noticed that there is a pronounced peak of stars at $\sim 250$\,Myr which we identify with the concentration of rapidly rotating young stars in the $UV$ diagram { noticed} by \citet{2004AN....325....3F}.  
This peak is significant at the $2\sigma$-level in the linear plot and significant in the logarithmic plot (Fig.~\ref{hlfig} and Table \ref{agesig}). While older intruders can be clearly separated from the Her-Lyr association, there are four to six younger stars whose membership cannot be conclusively determined. 

Measurements of $v\sin i$ were found for 13 additional stars. These measurements give an upper limit to the { gyrochronological} age. The full sample of 29 stars with { gyrochronological} age estimates is shown in gray in Fig.~\ref{hlfig}.
Including these additional members, the Her-Lyr association becomes even more pronounced. 

\begin{table}
\caption{Histogram statistics. }
\label{agesig}
\begin{center}
\begin{tabular}{lccc|c}
\hline\hline
number &$1\sigma$& $2\sigma$	& confidence	&max.\\
of stars&\multicolumn{2}{c}{per bin} & at 0.05 level&bin\\
\hline
\multicolumn{4}{c|}{\textit{linear}}&\\
only phot.	&	3 - 7	&	1 - 9	&	2 - 8	&11\\
all      	&	3 - 6	&	1 - 9	&	2 - 8	&18\\
\hline
\multicolumn{4}{c|}{\textit{logarithmic}}&\\
only phot.	&	2 - 5	&	1 - 7	&	1 - 6	&6\\
all       	&	3 -6	&	1 - 9	&	2 - 8	&8\\
\hline
\end{tabular}
\end{center}
\tablefoot{The table corresponds to Fig.~\ref{hlfig}. 
The number of stars per bin, predicted for a uniform distribution, is given for the upper gray data (linear  age, including upper limit data from $v\sin i$ measurements), the upper black data (without data obtained from $v\sin i$), lower gray data (same as upper gray data for logarithmic age), and lower black data (without $v\sin i$ data). The $0.05$-confidence interval for each bin is given in the fourth column and the maximal value is shown for comparison. The maximal value indicates the peak of the concentration of coeval stars that is identified with the Her-Lyr association.}
\end{table}

Most of the measurements indicate an association with an age of approximately$250$\,Myr while it is not clear whether the large spread is an intrinsic property of the association or whether it reflects the uncertainty of the { gyrochronology}. { In addition, the sample is contaminated by old intruders.} There are five stars that clearly show longer rotational periods and that are significantly older. From Table~\ref{hltab} it can be determined that these are the stars that also have {very} small lithium equivalent widths {or are even lithium-depleted}. While \object{V538\,Aur} and MN\,UMa
 show Li abundance and rotation {speed similar} to the Hyades, \object{V382\,Ser}, \object{39\, Tau}, and \object{LW\,Com} might be ordinary field stars, sharing the kinematics of Her-Lyr by chance.
The individual ages of stars in Table \ref{hltab} should be { considered} with care.

\subsection{Chromospheric activity}
\label{calcium}
Since the late-type Her-Lyr members are young and rotate rapidly, measurable chromospheric activity is expected. Chromospheric activity is commonly assessed from emission in the cores of the Ca\,II\,H\&K lines. The $\RHK$ index is corrected for photospheric flux contributions and allows one to compare the level of activity independent of spectral type. 

\begin{figure}
\centering
\resizebox{1.0\hsize}{!}{\includegraphics{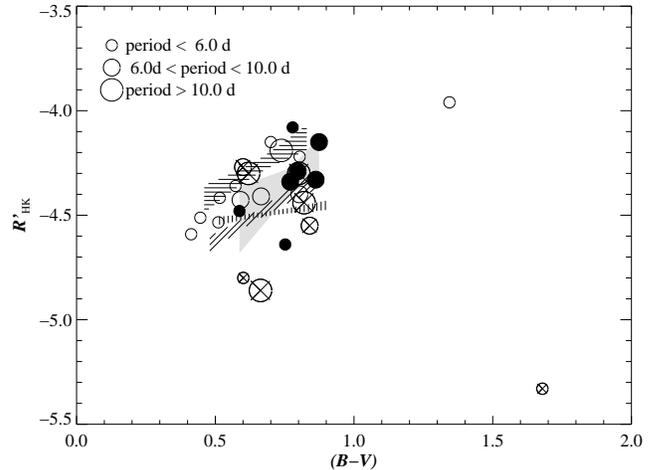}}
\caption{\label{plrhk} The activity level measured for Her-Lyr candidates is shown vs. color. The symbols exhibit individual measurements taken from the literature (Table~\ref{hltab}). { The symbols have the same meaning as in Fig.~\ref{lifig}}. {  The filled areas represent mean levels derived from the distributions of the Pleiades (horizontally hashed), Her-Lyr (gray), UMa (hashed), and the Hyades (vertically hashed). The widths are due to the uncertainty of the coefficients of the linear fits.}}
\end{figure}

Figure~\ref{plrhk} displays $\RHK$ measurements of Her-Lyr candidates taken from the literature (Table~\ref{hltab}) and compares to the activity levels of the Pleiades, the UMa group, and the Hyades \citep{2008ApJ...687.1264M}.

The distinction of the activity levels of the Pleiades, UMa, and the Hyades is not very clear as
the distributions of $\RHK$ strongly overlap. Therefore,
linear fits were obtained to get
the mean level of activity. The standard deviation of the fits qualitatively indicates the relative width of the distributions. Individual data points are only shown for the Her-Lyr stars which can be compared { to the average level of the Her-Lyr members according to Table \ref{memtab} (gray filled area)}. The difference of the average relations is less than the total widths of the distributions. { Fig.~\ref{plrhk} shows that for each association, the average levels of activity increase with color and are higher with younger average age of the respective association. Furthermore, the width of the distributions decrease with the average age of the respective association.}

The Her-Lyr relation is in between the Pleiades and the UMa relations regarding both the average activity level and the width supporting a young age of the Her-Lyr association between the age of Pleiades and UMa.

Among the kinematical Her-Lyr candidates, stars are identified above which display high levels of Li absorption and short rotation periods (Fig.~\ref{lifig}) unlike another group of purportedly older stars. The distinction is not as clear in terms of chromospheric activity. The older and younger objects widely overlap. This finding does not weaken the evidence of the two different groups identified since the activity distributions of the Pleiades, the Hyades, and the UMa group strongly overlap as well. 

We note, however, that the least active Her-Lyr candidates (HD\,4915 and HD\,207129) show activity levels similar to the less active Hyades members and are not among the rapid Li-rich rotators. these rotators are located above the mean Hyades relation, except for HD\,166 which also displays the least Li equivalent width among the { gyrochronological}ally young and Li-rich stars.

{ Similar to the gyrochronology, the { chromospheric activity} index can be used in a logarithmic fit to derive an age -- activity relation for main sequence stars. \citet{2008ApJ...687.1264M} derived 
\begin{equation}
\log t = -38.053 - 17.912\, \log\,R'_{\mathrm{HK}} -1.6675 \, \log(R'_{\mathrm{HK}})^2,
\end{equation}
where $t$ is the age in years and $R'_{\mathrm{HK}} $ is the {  chromospheric activity index}. The equation is appropriate between $-4.0 > R'_{\mathrm{HK}} > -5.1$. A similar histogram to the one in Fig.~\ref{hlfig} is depicted in Fig.~\ref{rhkage}. Treating \object{HD\,207129} and \object{G\,270-82} as old intruders, the evaluation of the logarithmic plot reveals an average age of approximately $285\,$ Myr for the subsample of Her-Lyr candidates with measured { activity level. The average age is in good agreement with that derived from the rotational period.}

\begin{figure}
\centering
\resizebox{1.0\hsize}{!}{\includegraphics{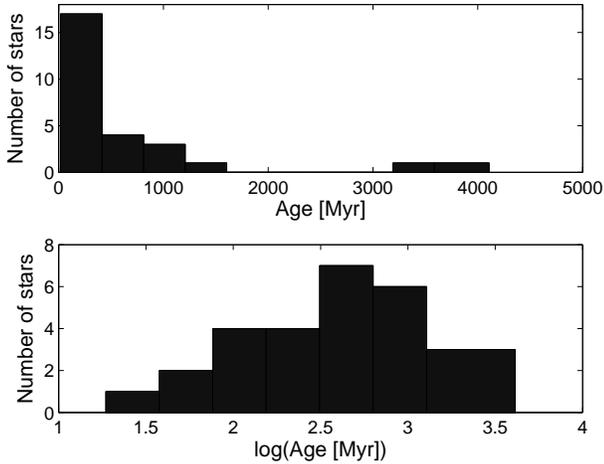}}
\caption{\label{rhkage} { The histogram of age derived from chromospheric activity is shown in linear and logarithmic scale as in to Fig.~\ref{hlfig}. The distribution shows that the majority of Her-Lyr candidates with measured activity level is younger than $500$\,Myr with the logarithmic plot showing a concentration around 250 Myr. The two older stars identified in the linear plot are \object{HD\,207129} and \object{G\,270-82}. The youngest star in the logarithmic plot is \object{DX\,Leo}.}}
\end{figure}
 }


\subsection{X-ray luminosity}
\label{xrays}
\begin{figure}
\centering
\resizebox{1.0\hsize}{!}{\includegraphics{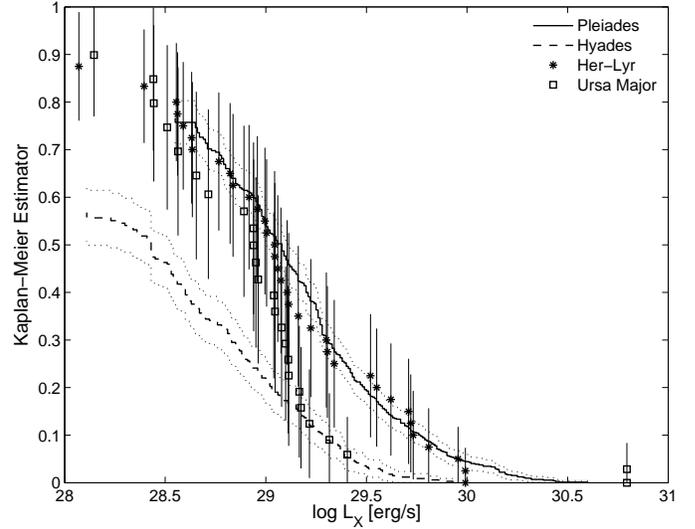}}
\caption{X-ray luminosity function of the Pleiades, Hyades, Her-Lyr (asterisks), and Ursa Major (squares). The logarithm of the luminosity measured by ROSAT is displayed with respect to the Kaplan-Meier estimator \citep{Kaplan1958,1985ApJ...293..192F} as an empirical cumulative distribution function of the X-ray luminosity. The X-ray data of the Pleiades and the Hyades was taken from the WEBDA database \citep{1996ASPC...90..475M}. For the Ursa Major and Her-Lyr stars, data from the \textit{ROSAT All-Sky Bright Source Catalog} \citep{1999A&A...349..389V} and the \textit{Second ROSAT PSPC Catalog} \citep{2000yCat.9030....0R} was used. Ursa Major members were taken { from the rather comprehensive study of} \citet{2003AJ....125.1980K}.
}
\label{xraylum}
\end{figure}

Young, late-type stars often show X-ray emission connected to coronal activity,  making the X-ray luminosity function a tool for age investigation. We convert count rate $R$ and hardness ratio from the \textit{ROSAT All-Sky Bright Source Catalog} \citep{1999A&A...349..389V} and the \textit{Second ROSAT PSPC Catalog} \citep{2000yCat.9030....0R}, if available, to X-ray fluxes for the Her-Lyr members and candidates { and for the} Ursa Major stars. The energy conversion factor of ROSAT is calculated by
\begin{equation}
\mathrm{ECF}= (5.3 \cdot \mathrm{HR1} + 8.31) \cdot 10^{-12}\,\frac{\mathrm{erg}}{\mathrm{cm}^2\cdot\mathrm{counts}},
\end{equation}
\citep{1995ApJ...450..392S} with the hardness ratio HR1. We obtained upper limits for undetected Her-Lyr and UMa stars from the \textit{ROSAT All Sky Survey} (RASS) data using \textit{XIMAGE}\footnote{XIMAGE is provided by the HEASARC X-ray archive.}. Upper limits on $R$ are converted to upper limits on luminosity $L_\mathrm{X}$ by a median ECF of $6\times 10^{-12}\,$erg$\,\cdot\,$cm$^{-2}\cdot$\,counts$^{-1}$. { The X-ray fluxes were transformed into luminosities via the stellar distances taken from \citet{2007A&A...474..653V}. For many stars in our sample, this was already done by other authors (e.g. \citealp{2003A&A...397..147P,2009A&A...499..129L}). We repeated the analysis for consistency.} In addition, we took X-ray luminosities and upper limits for the Pleiades and Hyades from the WEBDA database \citep{1996ASPC...90..475M}. We calculated the Kaplan-Meier estimator \citep{Kaplan1958} for right-censored data following the procedure outlined in \citet{1985ApJ...293..192F} to create an empirical cumulative distribution function. The result of this calculation is shown in Fig.~\ref{xraylum}.

From Fig.~\ref{xraylum} we can conclude that the Her-Lyr association is younger than the Hyades. Apparently, the X-ray luminosity is comparable to the Pleiades. The Ursa Major association includes a few very active stars, statistically comparable to the Hyades. On the low activity end, the Ursa Major association appears more Pleiades like. The Ursa Major group is therefore younger than the Hyades and older than the Pleiades and (as follows from the comparison of X-ray luminosity alone) also older than Her-Lyr.

\begin{table}
\caption{New X-ray upper limits for Her-Lyr and UMa stars. }
\begin{center}
\scriptsize
\begin{tabular}{lll|lll}
\hline\hline
Name & $\log R$ & $\log L_\mathrm{X}$&Name & $\log R$ & $\log L_\mathrm{X}$\\
          & [cts/s]                  & [erg/s]       &          & [cts/s]                  & [erg/s]\\
\hline
\multicolumn{3}{c|}{Her-Lyr}              		&HD 87696	& $<-11.934$ & $<29.046$\\
HD 4915 	& $<-12.406$ & $<28.338$		&HD 89025	& $<-12.002$ & $<29.925$\\
G 112-35 	& $<-13.146$ & $<27.259$		&HD 95418	& $<-12.218$ & $<28.637$\\
HD 25665	& $<-12.113$ & $<28.511$		&HD 102070	& $<-12.020$ & $<30.126$\\
EE Leo  	& $<-11.993$ & $<27.745$		&HD 106591	& $<-13.816$ & $<27.047$\\
NSV 6424	& $<-12.399$ & $<28.478$		&HD 109799	& $<-12.519$ & $<28.631$\\
$\alpha$ Cir& $<-13.423$ & $<27.094$		&HD 111397	& $<-11.739$ & $<30.514$\\
\multicolumn{3}{c|}{Ursa Major}          		&HD 112097	& $<-12.260$ & $<29.387$\\
HD 7804 	& $<-12.207$ & $<29.625$		&HD 112185	& $<-13.809$ & $<27.075$\\
HD 11171	& $<-11.484$ & $<29.324$		&GJ 519   	& $<-12.113$ & $<28.043$\\
HD 13959	& $<-11.913$ & $<29.276$		&HD 129246	& $<-11.796$ & $<29.745$\\
HD 18778	& $<-12.003$ & $<29.662$		&HD 129798	& $<-12.130$ & $<29.166$\\
HD 27820	& $<-12.265$ & $<29.981$		&HD 141003	& $<-12.190$ & $<29.242$\\
HD 40183	& $<-11.932$ & $<28.937$		&HD 173950	& $<-12.356$ & $<28.826$\\
HD 75605	& $<-12.308$ & $<29.443$		&HD 216627	& $<-11.808$ & $<29.655$\\
\hline
\end{tabular}
\end{center}
\tablefoot{$3\sigma$ upper limits on count rate and luminosity are given.}
\end{table}

\begin{figure}
\centering
\resizebox{1\hsize}{!}{\includegraphics[]{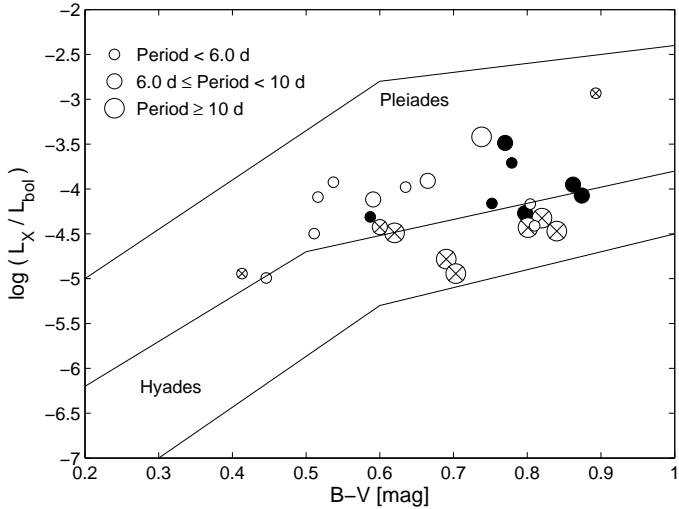}}
\caption{$B-V$ color index vs. the logarithmic ratio between X-ray and bolometric luminosity. Her-Lyr objects are plotted {  in the same way as in Fig.~\ref{lifig}}. Then, envelopes to the distributions of the Pleiades and Hyades are displayed according to  \citet[][Fig. 8]{2010A&A...521A..12M}. The figure shows that the age of Her-Lyr is in between the ages of the Pleiades and the Hyades and similar to the LA.
}
\label{lbol}
\end{figure}

{ Then, we also estimated the $L_{\rm X}$ over $L_{\rm bol}$
luminosity ratio for those of our objects {  for which} distance
and X-ray luminosity are known. We estimated the bolometric
luminosity as usual from V-band magnitude, bolometric
correction (from  spectral types following
\citealt{1995ApJS..101..117K}), and distance.
We then plotted the ratio of $L_{\rm X}$ to $L_{\rm bol}$
vs. the $B-V$ color in Fig.~\ref{lbol}.
We compare our objects with the Hyades and Pleiades
as plotted in \citet[Fig. 8]{2010A&A...521A..12M}.
Our Her-Lyr objects are located in between the Pleiades and Hyades
and are quite similar to the LA objects studied
in \citet{2010A&A...521A..12M}.
Hence, the age of Her-Lyr is in between the
ages of the Pleiades and the Hyades in agreement with our other approaches presented here. }

\subsection{The youth of the Her-Lyr association and canonical members}
\label{sect:youth}
It can be concluded that the age indicators give results which are in line with earlier estimates \citep{2004AN....325....3F,2006ApJ...643.1160L}. 
 
The seven stars (\object{V439\,And},
\object{EX\,Cet},
\object{EP\,Eri},
\object{DX\,Leo},
\object{HH\,Leo},
\object{PX\,Vir},
\object{HN\,Peg}) showing similar rotational periods, similar lithium equivalent width, and comparable space velocity can be used to define the properties of the Her-Lyr association. They are identified with the stars which make up the concentration of fast rotators in $UV$-space found by \citet{2004AN....325....3F} and studied by \citet{2006ApJ...643.1160L}. { They comprise three of the five stars which initially defined the Hercules moving group \citep{1998PASP..110.1259G}: V439 And, DX Leo, and HN Peg. The other two stars, MN UMa and NQ UMa, share the same space motion but have been discarded because they seem older than the other stars as is indicated by lithium equivalent width or { gyrochronological} age. Some objects that were adopted canonical members are doubted by  \citet{2004AN....325....3F}: EP Eri, DX Leo, and PX Vir. However, the evidence collected in this section that these three objects belong to the canonical members which form a physically associated group of stars is very convincing.}

\section{Kinematics}
\label{kine}
\begin{figure}
\centering
\resizebox{0.99\hsize}{!}{
\includegraphics[]{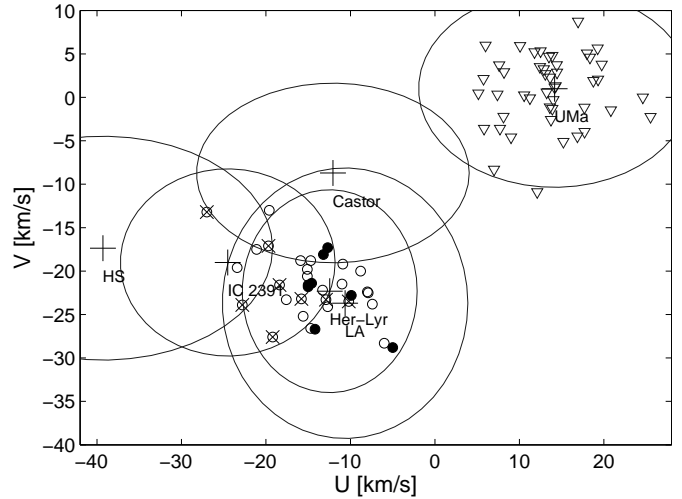}
}
\caption{The $UV$ diagram shows the distribution of nearby (within $25\,$pc) late-type stars in $UV$-space.
The Her-Lyr candidates are shown as black circles {as in Fig.~\ref{lifig}}. 
The data for Her-Lyr is taken from \citet{2006ApJ...643.1160L}.
Downward triangles are Ursa Major (UMa) stars.
The ellipses show the $3\sigma$ boundary of the displayed clusters as calculated from the space velocities of individual members.
$UV$ data are taken from \citet{2001MNRAS.328...45M}.
In addition, the large black pluses show the  mean velocities of some moving groups.
The location and the ellipse of the Her-Lyr association is calculated from the canonical members.
}
\label{hluvp}
\end{figure}

Using the canonical list of seven stars we can now redefine the average $UVW$ velocity of the association. { Space velocities are taken from \citet{2001MNRAS.328...45M}.} Figure~\ref{hluvp} displays the $VU$ plane of kinematic space. In addition to the Her-Lyr candidates, other associations are displayed. { We restrict this evaluation to stars within $25\,$pc to reach volume completeness.} The space velocity of Her-Lyr is similar to the \LA  so that there is controversy about whether the Her-Lyr association does exist as an individual entity or, rather if it is a part of the \LA \citep{2004AN....325....3F,2010A&A...521A..12M}. Taking the mean and the standard deviation of the seven canonical members, the space velocity of Her-Lyr  is
\begin{eqnarray}
\label{uvweq}
U&=&(-12.41 \pm 3.725)\mathrm{\,km\,s^{-1}}\\\nonumber
V&=&(-23.03 \pm 3.59)\mathrm{\,km\,s^{-1}}\\
W&=&(-8.11 \pm 3.80)\mathrm{\,km\,s^{-1}}.\nonumber
\end{eqnarray}
 
Based on the \citet{2004AN....325....3F} members, \citet{2006ApJ...643.1160L} derive $(U,V)=(-15.4,-23.4)\,\mathrm{km\ s}^{-1}$. With their refined member list, they get $(U,V)=(-13.19, -20.64)\,\mathrm{km\ s}^{-1}$ with a { standard} deviation of $(\sigma_U,\sigma_V)=(2.45,1.61)\,\mathrm{km\ s}^{-1}$. As they note, this deviation is smaller than that of other coeval MGs like Ursa Major. The dispersion in $W$ is $\sigma_W\approx3.4\,\mathrm{km\ s}^{-1}$ and compares well to our value. 

Except for \object{39\,Tau}, all rejected candidates concentrate in a specific region in the $UV$-diagram of Fig.~\ref{hluvp}. This could be a trace of an older stellar stream. The galactic { $V$ velocity} of this stream would then be in between that of Her-Lyr and IC\,2391 while the {  $U$ velocity} is similar to Her-Lyr.

\begin{table*}
\caption{\scriptsize \label{hltab} Spectroscopy and gyrochronology of all Her-Lyr candidates. {Canonical members are written in bold face.}}
{\scriptsize
\begin{center}
\begin{tabular}{llllcrl@{$\!$}rc@{$\,\!$}lr@{$\,\!$}cc|l}
\hline
\hline
&&&&\multicolumn{3}{c}{Spectroscopy}&\multicolumn{6}{c|}{Gyrochronology}&References\\
\hline
\multicolumn{2}{l}{Star}&SpT&$M_{\mathrm{tot}}$&$T_{\mathrm{eff}}$&$EW(\mathrm{Li})$&$R'_{\mathrm{HK}}$&$t_{R'_{\mathrm{HK}}}$&\multicolumn{2}{c}{$P$}&\multicolumn{2}{c}{$t_{\mathrm{gyro}}$}&$error$&\\
Name&HD/HIP &&$\Msun$&K&m\AA&&Myr&\multicolumn{2}{c}{days}&\multicolumn{2}{c}{Myr} &$\%$&\\
 \hline
{\bf ~1.~\object{V439\,And}}                    	&\object{HD\,166}		& K0\,V	& 0.99	&5471	&75.5	&-4.64&1440     	&          &5.69	&          &231	&12  	&2,3,4,5,9,10,13,23,25,30,32\\
 ~2.~\object{HIP\,1481}                     	&\object{HD\,1466}		& F9\,V	& 0.98      &6149	&125	&-4.36& 220    	&$\leq$&2.58	&$\leq$&161	&31  	&1,2,5,13,24,28\\
 ~3.~\object{G 270-82}                       	&\object{HD 4915}		& G1.5	& 0.91	&5630	&$\sim 0^{\#}$&-4.86&4180 &$\leq$	&22.8$^{\P}$	&$\leq$	&3409	&17	&7,20,23,24,25,27,30\\
 {\bf ~4.~\object{EX\,Cet}}                       	&\object{HD\,10008} 	& G5\,V	& 0.93	&5170	&103	&-4.29& 126    	&          &7.15	&          &315	&12  	&2,3,4,5,7,24,25,27\\
 ~5.~\object{LTT 10580}                      	&\object{HD\,10086} 	&G5\,IV	&1.04	&5685	&16  	&-4.60& 1430	&$\leq$&16.0	&$\leq$&1672	&15  	&7,13,20,23,24,30,32\\
~6.~\object{84\,Cet}                  		&\object{HD\,16765}	 	&F7\,IV	&1.20       &6240	&73  	&-4.534& 760	&$\leq$&1.8	&$\leq$&148	&77  	&3,9,11,13,24,33\\
~~~~7.~\object{84\,Cet\,B}                	&\object{HD\,16765\,B}	&K2\,V	&0.70	&\ldots	&\ldots	&\ldots& \ldots   &          &\ldots	&          &\ldots	&\ldots	&9\\	
~8.~\object{BC\,Ari} (SB)                          	&\object{HD\,17382} 	&K1\,V	&0.90	&5234	&$< 5$	& -4.44&  400  	&$\leq$&13.3	&$\leq$&901	&13  	&2,3,7,8,9,10,12,20,24,25\\
~~~~9.~\object{NLTT\,8996}               	&\object{HD\,17382\,C}	&M\,7 	&0.13	&\ldots	&\ldots  	&\ldots& \ldots	&        &\ldots	&          &\ldots	&\ldots	&7,9\\
{\bf 10.~\object{EP\,Eri}}                           	&\object{HD\,17925} 	& K1\,V	& 0.89	&5150	&212	&-4.33& 170	&           &6.725	&          &242	&11  	&1,3,4,5,6,8,9,10,11,16,23,25,30,31,33\\
11.~\object{1E\,0318-19.4}                	&\ldots                       	& K7\,V	& 0.65	&\ldots	&63  	&\ldots& \ldots	&           &\ldots	&          &\ldots	&\ldots 	&3,5,11,24\\
12.~\object{G\,112-35}                        	&\object{HIP\,37288}	& M0\,V	& 0.6	&4156	&\ldots	&\ldots& \ldots	&           &\ldots	&          &\ldots	&\ldots 	&3,5,24,25\\
13.~\object{HD\,25457}                         	&\object{HD\,25457}		& F5\,V	& 1.17	&6246	&118	&-4.417& 340 	&$\leq$ &3.35	&$\leq$&381	&59       	&3,4,5,8,9.13,15,20,24,25,27,29,33\\ 
14.~\object{G\,248-16}                      	&\object{HD\,25665}		&K2.5\,V  	&0.66	&4950	&\ldots	&\ldots& \ldots	&$\leq$&15.6	&$\leq$&719	&12  	&3,7,8,10,20,23,24,25,27,30,32\\
15.~\object{39\,Tau}	                         	&\object{HD\,25680} 	&G5\,V	&1.05	&5807	&61  	&-4.30& 140	&          &12.00 &         &1303	&17   	&2,7,8,9,23,24,25,30,32\\
16.~\object{V538\,Aur}                       	&\object{HD\,37394}		& K1\,V	& 0.91	&5200	& 2.2	&-4.55& 840	&          &10.00	&         &527	&12     	&2,3.4,5,7,8,16,23,24,25,30,32\\
~~~~17.~\object{LHS\,1775}                   	&\object{HD\,233153}	& M0.5\,V	& 0.45	&3740	&16  	&\ldots& \ldots  	&       &\ldots	&         &\ldots	&\ldots	&2,3,5,7,8,20,26\\
18.~\object{HD\,54371} (SB)                    	&\object{HD\,54371}		& G8\,V	& 1.03	&5503	&\ldots	&\ldots&\ldots	&          &\ldots	&         &\ldots	&\ldots  	&2,4,13,24,25,27,32\\ 
19.~\object{HD\,70573}                     	&\object{HD\,70573}		&G1/2\,V	& 1.0	&5899	&149	&\ldots&\ldots	&          &3.296	&         &125$^{*}$&14	&3,5,7,11,16,22,24\\
{\bf 20.~\object{DX\,Leo}}                         	&\object{HD\,82443}		& K1\,V	& 0.84	&5292	&176.3	&-4.08& 20	&         &5.38	&         &197	 &11	   	&2,3,4,5,6,7,8,9,16,24,25,32\\
{\bf ~~~~21.~\object{GJ\,354.1\,B}}          	&\ldots	         		& M5.5	& 0.1	&\ldots	&\ldots	&\ldots& \ldots	&         &\ldots	&         &\ldots   &\ldots	&2,9\\
22.~\object{EE\,Leo}	                         	&\object{HIP\,53020}	& M4	& 0.15	&\ldots	&\ldots	&-5.33&\ldots	&          &\ldots	&         &\ldots	 &\ldots 	&3,5,26\\
{\bf 23.~\object{HH\,Leo}}                       	&\object{HD\,96064}		& G5\,V	& 0.92	&5410	&114	&-4.34&190 	&         &6.9	&         &312	&12    	&2,3,4,5,9,24,25,33\\
{\bf ~~~~24.~\object{LTT\,4076}}             	&\ldots	         		& M0	& 0.6	&\ldots	&\ldots	&\ldots& \ldots	&         &\ldots	&         &\ldots 	&\ldots	&9\\
{\bf ~~~~25.~\object{BD-03\,3040C}}       	&\ldots 		         	& M0	& 0.6	&\ldots	&\ldots	&\ldots&\ldots	&          &\ldots	&         &\ldots	&\ldots	&9\\
26.~\object{MN\,UMa}                       	&\object{HD\,97334}		& G0\,V	& 1.08 	&5898	&10  	&-4.27&110 	&         &7.60$^{\dag}$&&643	&17  	&2,3,4,5,9,13,23,24,25,30,32\\
~~~~26a.~\object{Gl\,471 Ba}	               	&\ldots 	                  	& L4.5	&0.035	&\ldots	&\ldots	&\ldots&\ldots	&          &\ldots	&          &\ldots 	&\ldots	&9\\
~~~~26b.~\object{Gl\,471 Bb}	                	&\ldots 	                   	& L6 	&\ldots	&\ldots	&\ldots	&\ldots&\ldots	&          &\ldots	&          &\ldots 	&\ldots	&9\\
27.~\object{HR\,4758}                       	&\object{HD\,108799}	&G1/2	&1.04	&6009	&119	&-4.427&	370&$\leq$&8.24	&$\leq$&61$^{\S}$	&8	&7,12,13,24,33\\
~~~~28.~\object{GJ\,469.2\,B}                 	&\ldots                              	&K2/3	&0.7 	&4700	&\ldots	&\ldots& \ldots	&        &\ldots	&          &\ldots	&\ldots	&\\
29.~\object{LW\,Com}                       	&\object{HD\,111395}	& G5\,V	& 1.01	&5600	&6.3 	&-4.43& 370	&          &15.8	&          &1582	&16   	&3,4,5,13,16,20,23,24,25,27,30,32\\
30.~\object{HIP\,63317}                     	&\object{HD\,112733}	& G5\,V	& 0.93	&5512	&93.5	&-4.19& 50	&$\leq$&13.83	&$\leq$&1143	&14   	&5,8,11,13,24\\
~~~~31.~\object{HIP\,63322}                  	&\object{HIP\,63322}	& G6\,V	& 0.85	&4798	&112	&\ldots&\ldots	&          &\ldots	&          &\ldots	&\ldots	&,24\\
{\bf 32.~\object{PX\,Vir}} (SB)                   	&\object{HD\,113449}	& G5\,V	& 0.83	&5187	&142	&\ldots&\ldots	&          &6.47	&          &231	&11   	&2,3,4,5,6,7,9,24,25\\
33.~\object{NQ\,UMa}                       	&\object{HD\,116956}	& G9\,V	& 0.94	&5352	&40  	&-4.22& 70	&          &4.27	&          &124	&10   	&2,3,4,5,13,24,25,32\\
34.~\object{NSV\,6424}                     	&\object{HIP\,67092}	& K5  	& 0.65	&4162	&\ldots	&-3.96& \ldots	&          &\ldots	&          &\ldots	&\ldots  	&3,5,8,24\\
35.~\object{$\alpha$\,Cir}	                 	&\object{HD\,128898}	& A7\,V	& 1.81	&7452	&\ldots	&\ldots& \ldots	&          &4.47	&          &\ldots	&\ldots  	&13,15,21,24\\
~~~~36.~\object{LTT\,5826}                   	&\ldots	         		& K5\,V	& 0.72	&4346	&\ldots	&\ldots& \ldots	&          &\ldots	&          &\ldots	&\ldots 	&3,5\\
37.~\object{LTT\,6256}                     	&\object{HD\,139664}	& F4\,V	& 1.36	&7111	&$<9.1$	&-4.59& 1080	&$\leq$&0.94	&$\leq$&223	&94   	&3,13,14,18,24,29,33\\
38.~\object{LTT\,14623}                   	&\object{HD\,139777}	& G0   	& 1.03	&5746	&151	&-4.41& 320	&$\leq$&9.83	&$\leq$&767	&15   	&4,6,13,24,25,32\\
~~~~39.~\object{LTT\,14624}	               	&\object{HD\,139813}	& G5  	& 0.89	&5343	&142	&-4.40& 300	&$\leq$&8.33	&$\leq$&408	&12  	&2,4,20,23,24,25,30,32\\
40.~\object{V382\,Ser}                      	&\object{HD\,141272}	& G8\,V	& 0.88	&5270	&6     	&-4.30& 140	&1        &4.045	&          &1029	&13    	&2,3,4,6,9,20,24,25,32\\
~~~~41.~\object{HD\,141272}\,B              	&\ldots        			& M3	& 0.3	&\ldots	&\ldots	&\ldots& \ldots	&          &\ldots	&          &\ldots 	&\ldots  	&6,9\\
{\bf 42.~\object{HN\,Peg}}                        	&\object{HD\,206860}	& G0\,V	& 0.90	&5930	&124.8	&-4.48& 530	&          &4.70  	&          &296	&18   	&1,2,3,4,5,7,8,9,13,15,16,23,24,25,27,30,32\\
{\bf ~~~~42a.~\object{HN\,Peg}\,B}          	&\ldots        			& T2.5	& 0.021	&\ldots	&\ldots	&\ldots& \ldots	&          &\ldots	&          &\ldots 	&\ldots  	&9\\
43.~\object{HD\,207129}                   	&\object{HD\,207129}	& G0\,V	& 1.00	&6025	&44    	&-4.80&3200	&          &\ldots	&          &\ldots	&\ldots 	&1,3,5,13,14,19,23,24,28,30\\
44.~\object{V447 Lac}                       	&\object{HD 211472}	&K1\,V	&0.88      	&5258	&\ldots	&\ldots& \ldots	&          &4.43	&          &132	&10   	&2,3,7,9,24\\
45.~\object{LTT\,9081}                     	&\object{HD\,213845}	& F5\,V	& 1.33	&7037	&\ldots	&-4.512& 660	&$\leq$&2.18	&$\leq$&261	&29    	&3,5,8,13,15,24,25,29,33\\
46.~\object{V368\,Cep}                     	&\object{HD\,220140}	& G9\,V	& 0.9	&5144	&205.3	&-3.622& \ldots	&          &2.74	&          &49	&9    	&3,8,9,16,24,32\\
~~~~47.~\object{NLTT\,56532}       		&\ldots                 		& early M	& 0.2	&\ldots    	&\ldots	&\ldots& \ldots	&          &\ldots	&          &\ldots	&\ldots	&9\\
~~~~48.~\object{NLTT\,56725}               	&\ldots        			& M5.0	& 0.04	&\ldots	&\ldots	&\ldots& \ldots	&          &\ldots	&          &\ldots	&\ldots	&9\\
\hline
\end{tabular}\\
\end{center}
\tablebib{\scriptsize
(1)~\citet{1996AJ....111..439H}; (2)~\citet{1998PASP..110.1259G}; (3)~\citet{2001MNRAS.328...45M}; (4)~\citet{2004AN....325....3F}; (5)~\citet{2006ApJ...643.1160L}; (6)~\citet{2007AN....328..521E}; (7)~\citet{2008MNRAS.384..173F}; (8)~\citet{2010A&A...514A..97L}; (9)~\citet{2010ApJS..190....1R}; (10)~\citet{1992A&A...256..121T}; (11)~\citet{1995A&A...295..147F}; (12)~\citet{1996A&A...311..951F}; (13)~\citet{1999A&A...352..555A}; (14)~\citet{1999A&A...348..897L}; 
(15)~\citet{2003A&A...398..647R}; (16)~\citet{2003A&A...410..671M}; (17)~\citet{2003csss...12..881C}; (18)~\citet{2003A&A...409..251M}; (19)~\citet{2004A&A...414..601I}; (20)~\citet{2004ApJS..152..261W}; (21)~\citet{2005A&A...430.1143B}; (22)~\citet{2002AcA....52..397P,2003A&A...398..647R,2004AcA....54..153P,2005AcA....55...97P,2005AcA....55..275P}; (23)~\citet{2005ApJS..159..141V}; (24)~\citet{2006ApJ...638.1004A}; (25)~\citet{2010A&A...521A..12M}; (26)~\citet{2010AJ....139..504B}; (27)~\citet{2006A&A...450..735M}; (28)~\citet{2006A&A...460..695T}; (29)~\citet{2006A&A...446..267R}; (30)~\citet{2007ApJS..168..297T}; (31)~\citet{2007AJ....133.2524W}; (32)~\citet{2008A&A...489..923M}; (33)~\citet{2009A&A...493.1099S}; (34)~\citet{2010A&A...520A..79M}
}
\tablefoot{\scriptsize
The names in the first column represent the SIMBAD name, in the second column the HD or HIP name is given. Spectral type, mass, effective temperature, lithium equivalent width, calcium $HK$ index ${\RHK}$, and  { age derived from chromospheric activity} are presented as used in Sect.~\ref{youth}. The ages are calculated using the {  chromospheric activity index} and gyrochronology (this work). All other values are taken from the literature. Brown dwarf companions are further denoted by the primary star's number, followed by a small letter, starting with 'a'. Spectroscopic binaries have a "(SB)" after their name in the first column.  All companions are indented.\\
${\P}$: based on $v\sin i=13.97\,$km\,s$^{-1}$ \citep{2010A&A...520A..79M}, this object might be very young ($\lesssim 46\,$Myr); 
$*$: the double period is published as well, leading to an age of 425\,Myr;
$\dag$: a second period of 8.25 days is published, leading to an age of 740\,Myr;
$\S$: Possibly the result of a stellar merger \citep{2008MNRAS.384..173F};
$^\#$: Conversion from Li abundance to equivalent width \citep{1993AJ....106.1059S} reveals very small values, which are consistent with zero
}  }
\end{table*}

\begin{table*}
\caption{\scriptsize \label{hltab2} Photometric and astrometric properties of all Her-Lyr candidates.}
{\scriptsize
\begin{center}
\begin{tabular}{llrrr@{.}lrlr@{.}l@{$\ $}r@{.}l@{$\ $}r@{.}l|l}
\hline
\hline
&&&\multicolumn{5}{c}{Photometry}&\multicolumn{6}{c|}{Space velocity}&References\\
\hline
\multicolumn{2}{l}{Star}&$d_{\mathrm{p}}$&$B-V$&\multicolumn{2}{c}{$V$} & $V-H$&B.C. &\multicolumn{2}{c}{$U$}&\multicolumn{2}{c}{$V$}&\multicolumn{2}{c|}{$W$}&\\
Name&HD/HIP &pc& \multicolumn{5}{c}{phot. mag}&\multicolumn{6}{c|}{km\,s$^{-1}$}&\\
 \hline
{\bf ~1.~\object{V439\,And}}                     	&\object{HD\,166}     	&13.67	&0.752	&6&06 	&1.441&-0.20	&-15&0&-21&6&-10&0	&2,3,4,5,8,9,10,13,30,34\\
~2.~\object{HIP\,1481}                      	&\object{HD\,1466}   	&41.55 	&0.537 	&7&76	&1.212&-0.17	&  -8&8&-20&0&  -1&2 	&3,5,13,24,28,34\\
~3.~\object{G 270-82}                       	&\object{HD 4915}    	&21.52	&0.663	&6&98	&1.565&-0.21	&-15&6&-25&2&  -1&1	&7,24,30,34\\
{\bf ~4.~\object{EX\,Cet}}                 		&\object{HD\,10008} 	&23.95	&0.797	&7&66 	&1.761&-0.28	&-13&2&-18&1&-11&1	&2,3,4,5,7,24,34\\
~5.~\object{LTT 10580}                     	&\object{HD\,10086} 	&21.37 	&0.690	&6&60	&1.578&-0.22	&-19&7&-17&1&-19&4	&7,13,24,30,34\\ 
~6.~\object{84\,Cet}	                           	&\object{HD\,16765} 	&22.59	&0.511	&5&72	&1.078&-0.16	&-12&7&-24&1&  -2&7	&3,9,13,24,34\\
~~~~7.~\object{84\,Cet\,B}                 	&\object{HD\,16765\,B}	&21.62	&\ldots	&9&67	&\ldots&\ldots	&\multicolumn{6}{c|}{$3.3''$,\,$74.5$\,AU}	&9,34\\
~8.~\object{BC\,Ari} (SB)                          	&\object{HD\,17382} 	&24.64     	&0.820	&7&56 	&1.870&-0.31	&-22&8&-23&9&  -1&6	&2,3,7,8,9,10,24,34\\
~~~~9.~\object{NLTT\,8996}               	&\object{HD\,17382\,C}	&22.37	&\ldots	&15&59	&5.43&-3.29	&\multicolumn{6}{c|}{$20''$,\,$493$\,AU}   	&7,9,34\\
{\bf 10.~\object{EP\,Eri}}                           	&\object{HD\,17925} 	&10.35	&0.862     	&6&04 	&1.820&-0.37	&-15&0&-21&8&  -8&7	&3,4,5,6,9,10,24,30,34\\	
11.~\object{1E\,0318-19.4}                	&\ldots                       	&27.03	&1.139     	&10&24	&2.605&-0.55	&-12&7&-17&3&-11&8	&35,24,34\\
12.~\object{G\,112-35}                        	&\object{HIP\,37288}	&14.58	&1.379     	&9&65 	&3.538&-0.92	&-11&0&-21&5&-13&1	&3,5,24,34\\
13.~\object{HD\,25457}                         	&\object{HD\,25457}		&18.83    	&0.516     	&5&38 	&1.038&-0.16	&-6&0  &-28&3&-10&5	&3,4,5,8,9,13,24,31,34\\
14.~\object{G\,248-16}                      	&\object{HD\,25665}		&18.75	&0.952	&7&71    	&2.172&-0.48	&-7&4  &-23&8&-17&0	&3,7,8,10,24,30,34\\
15.~\object{39\,Tau}	                         	&\object{HD\,25680} 	&16.90	&0.620	&5&90	&1.402&-0.20	&-27& 0&-13&2& -7&1	&2,7,8,9,24,30,34\\	
16.~\object{V538\,Aur}                       	&\object{HD\,37394}		&12.28	&0.840    	&6&20	&2.219&-0.37	&-12&9&-23&3&-14&5	&2,3,4,5,7,8,24,30,34\\
~~~~17.~\object{LHS\,1775}                   	&\object{HD\,233153}	&12.44	&1.473	&9&75 	&3.817&-1.43	&-14&4&-22&9&-14&3	&2,3,5,7,8,34\\
18.~\object{HD\,54371} (SB)                     	&\object{HD\,54371}		&25.17	&0.700	& 7&08	&1.640&-0.23	&-21&1&-17&5&-15&7	&2,4,7,13,24,34\\
19.~\object{HD\,70573}                     	&\object{HD\,70573}		&40.36	&0.635	& 8&68    	&1.464&-0.22	&-14&7&-18&8&  -6&7	&3,5,24,31,34\\
{\bf 20.~\object{DX\,Leo}}                         	&\object{HD\,82443}		&17.79	&0.779	&7&06	&1.808&-0.28	&-9&9  &-22&8&  -5&6	&2,3,4,5,6,7,8,9,24,31,34\\
{\bf ~~~~21.~\object{GJ\,354.1\,B}}          	&\ldots	         		&\ldots	&\ldots	&14&7	&6.172&\-3.29	&\multicolumn{6}{c|}{$65''$,\,$1160\,$AU}	&2,9\\
22.~\object{EE\,Leo}	                         	&\object{HIP\,53020}	& 6.76	&1.679 	&11&64	&4.933&-2.56	&-7&9  &-22&5&-19&1	&3,5,34\\	
{\bf 23.~\object{HH\,Leo}}                        	&\object{HD\,96064}		&26.27	&0.770	&7&60  	&1.737&-0.28	&-14&2&-26&7&-0&6 	&2,3,4,5,9,24,34\\
{\bf ~~~~24.~\object{LTT\,4076}}             	&\ldots	         		&\ldots	&\ldots	&\multicolumn{2}{c}{\ldots}	&\ldots&\ldots	 &\multicolumn{6}{c|}{$11''$,\,$244\,$AU}		&9\\
{\bf ~~~~25.~\object{BD-03\,3040C}}       	&\ldots 		         	&\ldots	&\ldots	&\multicolumn{2}{c}{\ldots}	&\ldots&\ldots	 &\multicolumn{6}{c|}{$11''$,\,$244\,$AU}	 	&9\\
26.~\object{MN\,UMa}                       	&\object{HD\,97334}		&21.93	&0.600     	&6&40	&1.389&-0.19	&-15&8&-23&2&-11&2	&2,3,4,5,9,13,24,30,34\\
~~~~26a.~\object{Gl\,471 Ba}	           	&\ldots 	                  	&\ldots	&\ldots	&\multicolumn{2}{c}{\ldots}	&\ldots&\ldots	 &\multicolumn{6}{c|}{$90''$,\,$1974\,$AU}	&9\\
~~~~26b.~\object{Gl\,471 Bb}	           	&\ldots 	                   	&\ldots	&\ldots	&\multicolumn{2}{c}{\ldots}	&\ldots&\ldots	 &\multicolumn{6}{c|}{$90''$,\,$1974\,$AU}	&9\\
27.~\object{HR\,4758}                       	&\object{HD\,108799}	&24.65	&0.591 	&6&37     	&1.438&	&-23&4&-19&6&  -6&4	&7,13,24,31,34\\
~~~~28.~\object{GJ\,469.2\,B}                 	&\ldots                        	&24.65	&1.310	&9&58	&\ldots &-0.20	&\multicolumn{6}{c|}{$2$'',\,$49$\,AU}	&\\		
29.~\object{LW\,Com}                       	&\object{HD\,111395}	&16.93	&0.703	&6&29	&1.585&-0.82	&-18&4&-21&6&-9&2	&3,4,5,13,24,30,34\\
30.~\object{HIP\,63317}                     	&\object{HD\,112733}	&44.21	&0.738   	&8&65    	&1.720&-0.22	&-17&6&-23&3&-0&8	&5,8,13,24,34\\
~~~~31.~\object{HIP\,63322}                   	&\object{HIP\,63322}	&40.57	&0.852	&9&23	&2.284&-0.26	&-13&9&-20&3&-4&3	&24,31,34\\
{\bf 32.~\object{PX\,Vir}} (SB)                          	&\object{HD\,113449}	&21.69	&0.874	&7&70	&2.016&-0.37	&-5&0  &-28&8&-9&8	&2,3,4,5,6,7,9,24,31,34\\
33.~\object{NQ\,UMa}                       	&\object{HD\,116956}	&21.59	&0.804	&7&28	&1.809&-0.31	&-15&9&-18&8&-8&8	&2,3,4,5,13,24,34\\
34.~\object{NSV\,6424}                     	&\object{HIP\,67092}	&25.10	&1.344	&10&52	&3.212&-0.92	&-8&0  &-22&4&-1&8	&3,5,8,24,34\\
35.~\object{$\alpha$\,Cir}	                 	&\object{HD\,128898}	&16.57	&0.256	&3&17	&0.709&-0.10	&-10&9&-19&2&-10&8	&13,24,34\\
~~~~36.~\object{LTT\,5826}                   	&\ldots	         		&\ldots	&\ldots	&8&47	&2.780&-0.77	&\multicolumn{6}{c|}{$16''$,\,$262\,$AU}	&3,5\\
37.~\object{LTT\,6256}                     	&\object{HD\,139664}	&17.44	&0.413   	&4&64    	&0.908&-0.14	&-15&1&-19&8&-9&7	&3,5,13,14,24,34\\
38.~\object{LTT\,14623}                   	&\object{HD\,139777}	&21.85	&0.665     	&6&57     	&1.415&-0.20	&-14&7&-26&6&-2&2	&4,5,6,13,24,31,34\\	
~~~~39.~\object{LTT\,14624}	                	&\object{HD\,139813}	&21.51	&0.803     	&7&29     	&1.741&-0.29	&-14&7&-26&6&-2&2  	&2,4,5,24,30,31,34\\	
40.~\object{V382\,Ser}                      	&\object{HD\,141272}	&21.29	&0.801     	&7&44	&1.830&-0.31	&-19&2&-27&6&-14&0	&2,3,4,5,6,9,24,34\\
~~~~41.~\object{HD\,141272}\,B              	&\ldots        			&\ldots	&\ldots     	&\multicolumn{2}{c}{\ldots}	&\ldots&\ldots	 &\multicolumn{6}{c|}{$18''$,\,$383\,$AU}	&6,9\\
{\bf 42.~\object{HN\,Peg}}                        	&\object{HD\,206860}	&17.89	&0.587	&5&95	&1.362&-0.18	 &-14&6&-21&4&-11&0	&2,3,4,5,7,8,9,13,24,30,34\\
{\bf ~~~~42a.~\object{HN\,Peg}\,B}          	&\ldots        			&\ldots	&\ldots     	&\multicolumn{2}{c}{\ldots}	&\ldots&\ldots	 &\multicolumn{6}{c|}{$18''$,\,$383\,$AU}	&9\\
43.~\object{HD\,207129}                   	&\object{HD\,207129}	&15.99	&0.601	&5&57	&1.264&-0.19	 &-13&3&-22&2&-0&3	&3,5,13,14,24,28,30,34\\
44.~\object{V447 Lac}                       	&\object{HD 211472}	&21.54	&0.810	&7&50	&1.859&-0.31	 &-19&6 &-13&0&-5&9	&2,3,7,9,24,34\\
45.~\object{LTT\,9081}                     	&\object{HD\,213845}	&22.68	&0.446     	&5&21     	&0.944&-0.14	 &-15&1&-20&6&-12&9	&3,5,8,13,24,34\\
46.~\object{V368\,Cep}                     	&\object{HD\,220140}	&19.20	&0.893     	&7&53	&2.018&-0.42	 &-10&2&-23&5 &-5&5	&3,8,9,24,34\\
~~~~47.~\object{NLTT\,56532}       		&\ldots                 		&\ldots	&1.51	&12&24	&4.850&-1.64	 &\multicolumn{6}{c|}{$10''$,\,$192$\,AU}	&9\\
~~~~48.~\object{NLTT\,56725}               	&\ldots        			&19.38	&1.83 	&16&16	&6.318&-3.29	 &\multicolumn{6}{c|}{$16''$,\,$304$\,AU}	&9,34\\
\hline
\end{tabular}\\
\end{center}
\tablebib{\scriptsize
(2)~\citet{1998PASP..110.1259G}; (3)~\citet{2001MNRAS.328...45M}; (4)~\citet{2004AN....325....3F}; (5)~\citet{2006ApJ...643.1160L}; (6)~\citet{2007AN....328..521E}; (7)~\citet{2008MNRAS.384..173F}; (8)~\citet{2010A&A...514A..97L}; (9)~\citet{2010ApJS..190....1R}; (10)~\citet{1992A&A...256..121T}; (13)~\citet{1999A&A...352..555A}; (14)~\citet{1999A&A...348..897L}; (24)~\citet{2006ApJ...638.1004A}; (28)~\citet{2006A&A...460..695T}; (30)~\citet{2007ApJS..168..297T}; (34)~\citet{2007A&A...474..653V}
}
\tablefoot{\scriptsize
Object names follow the same conventions as in Table \ref{hltab}. Distance is given in the subsequent column, followed by the BVH photometry used in Sect.~\ref{photo}. { Hipparcos distances are available for most of the objects and have been adopted from \citet{2007A&A...474..653V}.} The $UVW$ velocity data occupies the next three columns. All values are taken from the literature. Separation and position angle of the companions are given in the $UVW$ columns. {The canonical members are written in bold face.}
} 
}
\end{table*}

\section{Notes on individual stars}
\label{indstar}
\paragraph{\object{84\,Cet} $=$ \object{HD\,16765}:} 
	The binary status of \object{84\,Cet} was discovered photographically by \citet{1975A&AS...21...51T}. The separation of $4\farcs01$ and a position angle of $306^{\circ}$ have both been measured by HIPPARCOS. However, \citet{2010ApJS..190....1R} published an updated value of { separation of} $3\farcs3$, yielding a projected distance of $74.5$\,AU between secondary and primary. A spectral type of K2V is estimated from photometry for the secondary. { Although this binary has been known for {  almost three decades}, the properties of the secondary are not well-known at all.} 

\paragraph{\object{BC\,Ari} $=$ \object{HD\,17382}:} 
This is a spectroscopic binary with an orbital solution giving an orbital period of $16.8\pm 2.2$\,yr \citep{2002AJ....124.1144L}. The mass function gives $M_\mathrm{A}=0.9\,\Msun$ and  $M_\mathrm{B}=0.19\,\Msun$. At an angular separation of $20''$, there is an additional M-dwarf companion (\object{NLTT 8996}) with a spectral type of M7 and a mass estimate of $\approx 0.13\,\Msun$ \citep{2007AJ....133..889L}. Including this wide companion, \object{BC\,Ari} is a triple system. 

\paragraph{\object{39\,Tau} $=$ \object{HD\,25665}:} 
A single epoch observation of a companion at $0\farcs216$ is reported \citep{1993AJ....106.1639M}, but this is contrasted with a constant radial velocity. 

\paragraph{\object{V538\,Aur} $=$ \object{HD\,37394}:}
	\object{V538\,Aur} is a K1 star at $12.2\,$pc. It is moving southward at $0\farcs5$ per year. At a distance of $98''$ and a position angle of $71^{\circ}$ \citep{2006PASP..118..218W}, the M0.5 star \object{LHS\,1775} (= HD\,233152) moves with the same proper motion (Fig.~\ref{visbin}a and \citealt{1956AJ.....61..405E}). It is surprising, however, that \citet{2006ApJ...643.1160L} classified \object{LHS\,1775} as a Her-Lyr member while \object{V538\,Aur} was excluded. Our analysis shows that according to lithium and rotation, this system is too old to be a Her-Lyr member. Adopting approximate { photometric} masses $M_\mathrm{A}=0.75\,M_{\odot}$ and $M_\mathrm{B}=0.45\,M_{\odot}$, assuming circular orbits, and neglecting projection effects, the system has a semi-major axis of $a\approx 1200$\,AU and an orbital period of $T\approx37\,000\,$yr.

\paragraph{\object{HD\,54371}:}
The photometric and kinematic analysis of \object{HD\,54371} suffers because this object is a spectroscopic binary with a period of $32.81$\,days \citep{1985PASP...97..355B,2010ApJS..190....1R}. The companion is a K dwarf with a magnitude difference of $\Delta m=2\,$mag or more.  The primary has a spectral type of G8 and {had previously been classified} a Her-Lyr candidate member even though the lithium feature is not that prominent \citep{2004AN....325....3F}.

\paragraph{\object{HD\,70573}:}
\cite{2007ApJ...660L.145S} reported { a planet candidate with a mass of} $m\sin i=6.1\pm 0.4\,M_\mathrm{Jup}$ at a semi-major axis of $1.76\pm0.05\,$AU and $\sim 850\,$days orbital period { using the radial velocity technique}. The planet has a comparably large eccentricity of $e=0.4\pm0.1$. It is one of the few radial velocity planets discovered in orbit around an apparently young star.

\paragraph{\object{DX\,Leo} $=$ \object{HD\,82443}:}
The K0 star \object{DX\,Leo} at $17.7\,$pc forms a common proper motion pair with the M5.5 star \object{GJ\,354.1\,B}  \citep{1998PASP..110.1259G}. Using POSS-I, POSS-II, 2MASS and additional Calar Alto 3.6\,m $\Omega$-Cass images we confirmed {the companion} \object{GJ\,354.1\,B} astrometrically and photometrically at a separation of $65''$ and a position angle of $67^{\circ}$ (Fig.~\ref{visbin}b). Using the photometric masses $M_\mathrm{A}=0.84\,M_{\odot}$ and $M_\mathrm{B}=0.21\,M_{\odot}$ this translates into a linear separation of $a\approx1160\,$AU and a period of $T\approx38\,400\,$yr ($e=0$ assumed).


\paragraph{\object{HH\,Leo} $=$ \object{HD\,96064}:}
The multiplicity of \object{HH\,Leo} was investigated by \citet{1980ApJS...44..111H}. The object, located at a separation of 11'' ($a\approx244\,$AU, $T\approx3378$\,yr) was later found to be a binary \citep{1993AJ....106..352H}, i.e. the whole system is a hierarchical triple (Fig.~\ref{hhleo}). 

\paragraph{MN\,UMa $=$ \object{HD\,97334}:}
At a separation of $90''$ at $245^{\circ}$, \citet{2001AJ....121.3235K} discovered a L4.5 companion (GJ\,417\,B). Common proper motion was proven using the 2MASS catalog. This object is too red and too faint to be visible {on} Schmidt plates so we missed it in our analysis. Later, \citet{2003AJ....126.1526B} {resolved} this brown dwarf companion into a binary using HST WFPC2. They report a separation of $70\,$mas or $1.9$\,AU and a magnitude difference of $\sim1$  (see also \citealp{2010ApJS..190....1R}). 

\paragraph{\object{HR\,4758} $=$ \object{HD\,108799}:} This object might be an example of a former stellar merger \citep{2008MNRAS.384..173F}. This idea arises from the fact that this star has all indicators of a young object (rotation, lithium, H$\alpha$, etc.) but looks like an old object on evolutionary tracks.  From its kinematics, { however,} a Her-Lyr membership cannot be excluded. It has a K-dwarf companion with $0.7\,\Msun$ at a separation of $\sim2''$ and the orbital period is estimated to be $P=151$\,yr \citep{2008MNRAS.384..173F}.

\paragraph{\object{HIP\,63317} $=$ \object{HD\,112733}:}
The G5 star \object{HD\,112733} forms a common proper motion pair with its neighboring twin, the G6 star \object{HIP\,63322}. \citet{1986A&AS...66..131H} discovered this pair with a projected separation of $a=807\,$AU giving a period of $T=17\,124\,$yr if the projected separation is close to the semi-major axis. Given the differences in parallax and proper motion which are slightly larger than the corresponding uncertainty, it is not clear however whether these two stars are bound.

\paragraph{\object{PX\,Vir} $=$ \object{HD\,113449}:}
This is a stellar binary. Various photometric and spectroscopic measurements were taken \citep{1950ApJ...112...48M,2000AJ....120.1006G,2010AAS...21531702L}. A photometric period of $231$\,days and radial velocity changes of $20$\,km\,s$^{-1}$  were observed. The companion was resolved in 2007 with {a brightness difference of} $\Delta H\sim1.6$\,mag, {a} separation of $35.65\pm0.6$\,mas, and {a} position angle of $225.2^{\circ}\pm0.2^{\circ}$ \citep[][]{2010ApJS..190....1R}. Astrometric and spectroscopic measurements are consistent.

\paragraph{\object{$\alpha$\,Cir} $=$ \object{HD\,128898}:}
$\alpha$ Cir is the brightest star in the Circinius constellation. Being a nearby late A-type star, it is too bright for analysis via Schmidt plates. In addition, it shows astroseismic variations as studied in detail by \citet{2009MNRAS.396.1189B}. The K5 star \object{LTT\,5826} is located at the same distance, {thus it} is probably a wide companion {of} $\alpha$ Cir. This companion candidate was considered a Her-Lyr candidate by \citet{2006ApJ...643.1160L} while \object{$\alpha$\,Cir} remains unattended in their work.

\paragraph{\object{LTT\,14623} $=$ \object{HD\,139777} and \object{LTT\,14624} = \object{HD\,139813}:}
The well-known common proper motion pair was first mentioned by \citet{1960AJ.....65...60S}. With spectral types of G0 and G5, both are solar analogs. The upper limits on the age derived via $v\sin i$ measurements are, however, insufficient to confirm {a probable Her-Lyr membership}. Currently, their linear separation is $\sim680\,$AU {corresponding to an orbital period of} $T\approx 13\,000\,$yr in case of a circular orbit.

\paragraph{\object{V382\,Ser} $=$ \object{HD\,141272}:}
\citet{2007AN....328..521E} confirmed that the object $\approx18''$ north of \object{V382\,Ser} is a co-moving companion using Schmidt plates, $\Omega$-Cass images (Fig.~\ref{visbin}c), the 2MASS catalog as well as EMMI/NTT spectroscopy. {Spectroscopy indicates} a spectral type of M3 and a  mass of $\sim 0.3\,M_{\odot}$ \citep{2007AN....328..521E}  for the companion \citep[see also][]{2010ApJS..190....1R}.

\paragraph{\object{HN\,Peg} $=$ \object{HD\,206860}:}
\citet{2007ApJ...654..570L} discovered a T2.5 dwarf companion using Spitzer IRAC and confirmed the common proper motion using the 2MASS catalog. In the archival Schmidt plates, the object is not detected. However, we imaged the companion with ALFA (see Fig.~\ref{HNPeg}). At a mass of $0.021\,M_{\odot}$ \citep{2007ApJ...654..570L}, this is the least massive directly detected member of the Her-Lyr association \citep[see also][]{2010ApJS..190....1R}.

\paragraph{\object{V368\,Cep} $=$ \object{HD\,220140}:}
This G9V star is probably the youngest {Her-Lyr candidate in our list}. Since it was not in the initial sample of \citet{2004AN....325....3F}, it was also not considered by \citet{2006ApJ...643.1160L}. The kinematics make it a good candidate for the Her-Lyr association although new lithium equivalent widths \citep{2010A&A...514A..97L} and photometry of the star and its companions suggest a somewhat younger age (12-20\,Myr, \citealp[see][]{2010ApJS..190....1R} and \citealp{2007ApJ...668L.155M}). The gyrochronology suggests $\sim 50\,$Myr. 
{ The celestial location co-incides with the Cepheus-Cassiopeia complex and a group of four co-moving isolated T\,Tauri stars identified by \citet{2010A&A...520A..94G}. They argue, however, that the proper motion and the distances rule out a common origin with V368 Cep and its companions.} The brighter companion, NLTT\,5632, at a separation of $10''$, shows common proper motion with the primary and its photometry yields a spectral type of early M at the primary's distance \citep{2010ApJS..190....1R}. The second companion, \object{NLTT\,56725}, at $16''$ separation is a late-M dwarf which was discovered by \citet{2005AJ....129.1483L} and later confirmed by measuring the parallax \citep{2007ApJ...668L.155M}. This star is overluminous in the $K_\mathrm{s}$ band, again indicating a young age of the system.

All directly detected companions outside $10\,$AU are listed separately in Tables \ref{hltab}, \ref{hltab2}, and \ref{memtab}.

\section{Discussion}
\label{concl} 
Tables~\ref{hltab} and \ref{hltab2} summarize all properties of Her-Lyr candidates used in this analysis. In the following, our conclusions regarding age, membership, and multiplicity statistics are presented. Quantitative membership criteria are developed as far as possible. 


\citet{2006ApJ...643.1160L} already noted that isochrone fitting fails to give tight constraints on the age of the association. The stars are simply too old and most of them have reached the main-sequence.
We have searched for late-type companions to Her-Lyr candidates to obtain some improvement on that issue. Late-M stars may still be contracting and provide a possibility of deriving an association age via theoretical isochrones. However, our analy\-sis shows that either the models themselves or the { calibration of observables} (color and magnitude) lacks precision at the late-type end. We suspect that the problems occur because M dwarfs become fully convective around spectral type M3-M4. The isochrone method fails to derive an age estimate for the Her-Lyr association. Alternatively, the late-M stars in question (especially the companions to V368\,Cep) may indeed be very young.

%
From the list of candidates, selected via space velocity, we were able to identify seven canonical members using gyrochronology and lithium equivalent width. These members define the properties of the association hence the average { gyrochronological} age of the Her-Lyr association is (with $1\sigma$ error bars)
\begin{equation}
\label{ageeq}
\mathrm{age} =257 \pm 46 \mathrm{\ Myr}.
\end{equation}

The youth of the Her-Lyr association is confirmed by the X-ray luminosity function, { the $L_X/L_{\mathrm{bol}}$ ratio}, the analysis of lithium equivalent width, the $R'_{\mathrm{HK}}$ index, { and the age derived from chromospheric activity}. { The activity--age relation used here ignores spectral type or color hence should be used with care when looking at individual ages. The average overall candidates ($\sim 285\,$Myr), however, coincides with the { gyrochronological} age.} The Her-Lyr association is at least as old as the Pleiades and younger than the Hyades. There are some hints that an age similar to Ursa Major can be assumed. 

%
The canonical member list is identified in Sect.~\ref{sect:youth} and is composed of the young lithium-enriched stars in Fig.~\ref{lifig}, i.e., 
V\,439\,And (\object{HD\,166}), 
\object{EX\,Cet} (\object{HD\,10008}),
\object{EP\,Eri} (\object{HD\,17925}), 
\object{DX\,Leo} (\object{HD\,82443}), 
\object{HH\,Leo} (\object{HD\,96064}), 
\object{PX\,Vir} (\object{HD\,113449}), and
\object{HN\,Peg} (\object{HD\,206860}). Several additional stars
(\object{G\,112-35}, \object{HD\,25457}, \object{HIP\,63317} \& \object{HIP\,63322}, \object{NSV\,6424}, \object{$\alpha$\,Cir} A\&B, \object{LTT\,6256}, \object{LTT\,14623} \& \object{LTT\,14624}, and \object{LTT\,9081}) show good agreement, but cannot be used for the canonical list of members since some information is missing (typically lithium measurements or rotational periods).

Based on the average properties derived from the canonical list of members, three membership criteria are defined in order to establish membership of additional candidates:
\begin{enumerate}
\item The $U$ and $V$ velocities must not deviate by more than $2\,\sigma$ from the mean $U$ and $V$ velocities of the association (Eq.~\ref{uvweq}).
\item 
The lithium equivalent width of a Her-Lyr candidate should be as least as high as for an UMa member of the same effective temperature (applicable above 5000K).
\item { Chromospheric activity should not be significantly below the mean levels of the Hyades and UMa in Fig.~\ref{plrhk}.} As the calcium $\RHK$ index is a very uncertain criterion, we use it as a supporting argument. 
\item The { gyrochronological} age should be within $2\,\sigma$ of the mean { gyrochronological} age of the association (Eq.~\ref{ageeq}).
\end{enumerate}

{ These criteria are applied to each candidate. The results are given in Table \ref{plrhk}.} Given our definition of Her-Lyr like kinematics (Eq. \ref{uvweq}) and age (Eq. \ref{ageeq}), we give the deviation (in $\sigma$) for each candidate in Table~\ref{memtab}. We consider a star a doubtful candidate if it has a deviation of more than two $\sigma$ in one or more properties and a non-member in the case of a deviation of more than three $\sigma$. { The lithium equivalent width and the chromospheric activity index $\RHK$ are used as an additional qualitative argument.}

The member list also includes all co-moving companions of {  members} which are \object{GJ\,354.1\,B}, \object{LTT\,4076}, BD-03\,3040\,C, \object{PX\,Vir}\,B, and \object{HN\,Peg}\,B { (see Sect.~\ref{indstar})};  \object{HIP\,63317} (\object{HD\,112733}) with its companion \object{HIP\,63322}  are possible members  { as well as \object{1E\,0318-19.4}, \object{G\,248-16}, and \object{HD\,54371}.} 
{ The criteria on Li and $\RHK$ only apply to late-type stars. For this reason, \object{LTT 6256} has been kept as candidate even though no significant Li absorption has been measured.

Interestingly, all doubtful candidates have a similar age too young to include them in our current list of Her-Lyr members. The question arises whether this is another group of stars or if the age distribution is bimodal indicating that star formation did not occur as a single event in the Her-Lyr association. At present, the number of stars and the data available are insufficient to firmly answer this question.}

\begin{table}
\caption{\scriptsize Membership of Her-Lyr candidates. 
}
\label{memtab}
\begin{center}
\scriptsize
\begin{tabular}{l@{$\quad$}c@{$\quad$}c@{$\quad$}c@{$\quad$}c@{$\quad$}c@{$\quad$}c@{$\quad$}c@{$\quad$}c}
\hline
\hline
Primary			&	$\sigma_{UV}$&	$\sigma_{W}$	&	EW(Li)	&	$R'_{\mathrm{HK}}$	&	$\sigma_{\mathrm{age}}$	&	Mem. &	 F04	& L06\\
\hline
{\bf ~1.~\object{V439\,And}}	&	1	&	1	&$\checked$      	&$\checked$      	&	1		&	+	&	+	&	+	\\
~2.~\object{HIP\,1481}		&	1	&	1.9	&$\checked$       	&$\checked$     	&$\gtrsim$2.2	&	o	&		&	o	\\
~3.~\object{G\,270-82}		&	1	&	1.9	&	?       		&$\mbox{\lightning}$&$\lesssim$67.8&	-	&		&		\\
{\bf ~4.~\object{EX\,Cet}}		&	1.4	&	1	&$\checked$       	&$\checked$     	&	1.2		&	+	&	+	&	+	\\
~5.~\object{LTT\,10580}		&	2.0	&	3.0	&$\mbox{\lightning}$&$\mbox{\lightning}$&$\lesssim$30.4&	-	&		&		\\
~6.~\object{84\,Cet}			&	1	&	1.5	&$\checked$      	&$\checked$     	&$\gtrsim$2.5	&	o	&		&		\\
~~~~~7.~\object{84\,Cet\,B}		&	1	&	1.5	&	?       		&	?                	&	?		&	o	&		&		\\
~8.~\object{BC\,Ari}			&	2.8	&	1.8	&$\mbox{\lightning}$&$\checked$     	&$\lesssim$13.8&	-	&		&		\\
~~~~~9.~\object{NLT\,8996}		&	2.8	&	1.8	&	?        		&	?        		&	?		&	-	&		&		\\
{\bf 10.~\object{EP\,Eri}}		&	1	&	1	&$\checked$      	&$\checked$      	&	1		&	+	&	o	&	o	\\
11.~\object{1E\,0318-19.4}	&	1.6	&	1	&$\checked$     	&	?	         	&	?		&	?	&		&	o	\\
12.~\object{G\,112-35}		&	1	&	1.4	&	?         		&	?         		&	?		&	?	&		&	+	\\
13.~\object{HD\,25457}		&	1.8	&	1	&$\checked$     	&$\checked$           	&$\lesssim$2.6	&	?	&	o	&	-	\\
14.~\object{G\,248-16}		&	1	&	1.7	&	?                	&	?                	&$\lesssim$9.9	&	?	&		&		\\
15.~\object{39\,Tau}			&	2.4	&	2.0	&$\checked$     	&$\checked$     	&	22.5		&	-	&		&		\\
16.~\object{V538\,Aur}		&	1	&	1.7	&$\mbox{\lightning}$&$\checked$     	&	5.8     	&	-	&	+	&	o	\\
~~~~17.~\object{LHS\,1775}		&	1	&	1.7	&$\mbox{\lightning}$&	?                	&	?       	&	-	&		&	+	\\
18.~\object{HD\,54371}		&	2.4	&	2.0	&	?        		&	?        		&	?		&	?	&	o	&		\\
19.~\object{HD\,70573} 		&	1.2	&	1	&$\checked$		&	?        		&	3.0		&	o	&		&	+	\\
{\bf 20.~\object{DX\,Leo}}		&	1	&	1	&$\checked$     	&$\checked$		&	1.4		&	+	&	o	&	o	\\
~~~~{\bf 21.~\object{GJ\,354.1\,B}}	&	1	&	1	&	?	         	&	?	         	&	?		&	+	&		&		\\
22.~\object{EE\,Leo}			&	1.3	&	2.9	&	?       		&	o	         	&	-		&	-	&		&	+	\\
{\bf 23.~\object{HH\,Leo}}		&	1.1	&	2.0	&$\checked$     	&$\checked$		&	1.2		&	+	&	+	&	o	\\
~~~~{\bf 24.~\object{LTT\,4076}}		&	1.1	&	2	&	?       		&	?       		&	?		&	+	&		&		\\
~~~~{\bf 25.~BD\,033040\,C}			&	1.1	&	2	&	?	         	&	?       		&	?		&	+	&		&		\\
26.~\object{MN\,UMa}		&	1	&	1	&$\mbox{\lightning}$&$\checked$     	&	8.3      	&	-	&	+	&	o	\\
27.~\object{HR\,4758}		&	3.0	&	1	&$\checked$     	&$\checked$     	&$\gtrsim$4.3	&	o	&		&		\\
~~~~28.~{\bf \object{GJ\,469.2\,B}}		&	3.0	&	1	&	?	         	&	?	         	&	?		&	o	&		&		\\
29.~\object{LW\,Com}		&	1.7	&	1	&$\mbox{\lightning}$&$\checked$     	&	28.5     	&	-	&	o	&	o	\\
30.~\object{HIP\,63317}	    	&	1.4	&	2.0	&$\checked$     	&$\checked$     	&$\lesssim$19.0&	?	&		&	o	\\
~~~~31.~\object{HIP\,63322}		&	1.4	&	2.0	&$\checked$     	&	?	         	&	?		&	?	&		&	o	\\
{\bf 32.~\object{PX\,Vir}}	    	&	2.0	&	1	&$\checked$     	&	?	         	&	1		&	+	&	o	&	-	\\
33.~\object{NQ\,UMa}		&	1.2	&	1	&	o                 	&$\checked$     	&	3.0    	&	o	&	+	&	o	\\
34.~\object{NSV\,6424}		&	1.2	&	1.7	&	?	         	&$\checked$     	&	?		&	?	&		&	o	\\
35.~\object{$\alpha$\,Cir}		&	1.1	&	1	&	?       		&	?	         	&	?		&	?	&		&		\\
~~~~36.~\object{LTT\,5826}		&	?	&	?	&	?       		&	?	         	&	?		&	?	&		&	o	\\
37.~\object{LTT\,6256}		&	1	&	1	& ? &$\checked$      	&$\gtrsim$1	&	?	&	+	&	+	\\
38.~\object{LTT\,14623}		&	1	&	1.6	&$\checked$      	&$\checked$      	&$\lesssim$11.0&	?	&	+	&	o	\\
~~~~39.~\object{LTT\,14624}		&	1	&	1.6	&$\checked$      	&$\checked$       	&$\lesssim$3.2	&	?	&	+	&	o	\\
40.~\object{V382\,Ser}		&	1.9	&	1.6	&$\mbox{\lightning}$&$\checked$      	&	16.6  	&	-	&	+	&	o	\\
~~~~41.~\object{HD\,141272}\,B	&	1.9	&	1.6	&	?       		&	?       		&	?		&	-	&		&		\\
{\bf 42.~\object{HN\,Peg}}		&	1	&	1	&$\checked$     	&$\checked$     	&	1		&	+	&	+	&	+	\\
43.~\object{HD\,207129}		&	1	&	2.1	&$\checked$       	&$\mbox{\lightning}$&	?		&	-	&		&	o	\\
44.~\object{V447\,Lac}		&	2.8	&	1	&	?       		&	?	         	&	2.8		&	o	&		&		\\
45.~\object{LTT\,9081}		&	1	&	1.3	&	?       		&$\checked$     	&$\gtrsim$1	&	?	&		&	+	\\
46.~\object{V368\,Cep}		&	1	&	1	&$\checked$       	&$\mbox{\lightning}$&	4.6		&	-	&	-	&		\\
~~~~47.~\object{NLTT\,56532}	&	1	&	1	&	?       		&	?       		&	?		&	-	&		&		\\
~~~~48.~\object{NLTT\,56725}	&	1	&	1	&	?       		&	?       		&	?		&	-	&		&		\\
\hline
\end{tabular}
\end{center}
\tablefoot{\scriptsize { The numbers in Cols. 2 and 3 give the deviation of the space velocity of each star from the mean space velocity of Her-Lyr in Sect. \ref{kine} in units of $\sigma$. The criteria defined in Sect. \ref{kine} are applied to each single candidate. A discrepancy of more than two $\sigma$ in $UV$ velocity and age disqualifies a star as a probable member.} For lithium, a more qualitative argument ($\mathrm{EW(Li)_{UMa}< EW(Li)_{Her-Lyr}\lesssim EW(Li)_{Pleiades}}$) is given. A Her-Lyr candidate is assigned a $\checked$ if the equivalent width fits that of the canonical members, a $\mbox{\lightning}$ if the star is lithium-depleted, and a ? if no information was found { or if the criterion is not applicable to this particular object. Similarly, a $\checked$ is given if a candidate displays significant chromospheric activity $\RHK$, otherwise $\mbox{\lightning}$ or ? if unknown.} The $\gtrsim$ and $\lesssim$ signs in Col. 6 indicate that the age was derived based on an upper limit of the rotational period, i.e., $v\sin i$. Based on these criteria, the membership probability is qualitatively indicated by a + for a high membership probability and a o for doubtful membership. Improbable members do not fulfill at least one of the criteria and are indicated by -. In case of unavailable information or upper limits, the candidate is kept as a possible member and marked by ?  {( Col. 7)}. { For comparison, results by \citet{2004AN....325....3F} and \citet{2006ApJ...643.1160L} are given in the last two columns.} {In cases of resolved multiple systems where common proper motion is confirmed (see Sects. \ref{multi} and \ref{indstar}), similar $UVW$ velocity is assumed and the membership is assessed as for the primary. { Canonical members are written in bold face.}}
}
\end{table}

A critical review of earlier assessments must neccesarily be done before reconsidering the member list. Therefore, the member lists of \citet{2004AN....325....3F} and \citet{2006ApJ...643.1160L} are given for comparison in the last two columns of Table~\ref{memtab}. \citet{2004AN....325....3F}  identified a large initial list of candidates. \citet{2006ApJ...643.1160L} discarded \object{HD\,25457} since the kinematics of that star match the B4 subgroup of the Pleiades. The systems \object{HH\,Leo} ($=$\object{HD\,96064}), \object{NSV\,6424} ($=$\object{HIP\,67092}), \object{HD\,207129}, and the binaries \object{HIP\,63317} ($=$\object{HD\,112773}) \& \object{HIP\,63322}, and \object{LTT\,14623} ($=$\object{HD\,139777}) \& \object{LTT\,14624} ($=$\object{HD\,139813}) were {  doubted} because of their $W$ velocity. This is not a strong argument, however, because the Her-Lyr candidates have already completed $\sim 1$ galactic orbit so that the $W$ velocity can be significantly changed thanks to the galactic potential \citep[see e.g.,][Fig.~23 and discussion]{2004AN....325....3F}. 

The \object{PX\,Vir} system is discarded by \citet{2006ApJ...643.1160L} because it was earlier classified as an AB\,Dor member \citep{2004ApJ...613L..65Z}. It is not clear, however, {to what extent} the Her-Lyr association differs from the AB\,Dor moving group.
In contrast to \citet{2006ApJ...643.1160L}, we exclude stars depleted of lithium and showing excessively slow rotational behavior. {  Seven} stars (\object{HIP\,1481}, { \object{84\,Cet}}, \object{HD\,70573}, { \object{HR\,4758}}, \object{NQ\,UMa}, { \object{V447\,Lac}}, and \object{HD\,220140}) are younger than the average age of the Her-Lyr association.
 
The present work adds new membership criteria to the kinematic criteria in a coherent way, based on a well-defined list of canonical members and consolidates the properties and member list of the Her-Lyr association.

The multiplicity study revealed that our sample consists of 35 stellar systems, counting just the primaries, including the spectroscopic binaries \object{BC\,Ari}, \object{HD\,54371}, and \object{PX\,Vir} ($=$\object{HD\,113449}). Fithteen ($42.9\%$) of these are multiple, including four ($11.4\%$) triple systems and eleven ($31.4\%$) binaries. Two stars, \object{MN\,UMa} and \object{HN\,Peg} have brown dwarf companions. Comparing this with the $25\,$pc volume-complete sample of \citet{2010ApJS..190....1R}, our results are consistent. However, the the small quantity of statistical data should be used with care. We note that because of the extension of the sample by \citet{2006ApJ...643.1160L}, stars outside $25\,$pc\footnote{With the re-reduction of the {\it Hipparcos} catalog by \citet{2007A&A...474..653V}, some of the stars, included in the 25\,pc sample of \citet{2004AN....325....3F}, are now located outside the 25\,pc sphere (e.g., \object{HH\,Leo}). This necessarily leads to inconsistencies regarding the 25\,pc sample.} are considered, as well as non-solar like stars. \citet{2007ApJ...660L.145S} detected a planet candidate around \object{HD\,70573}. To our knowledge no further planetary mass companion was found for any Her-Lyr member or candidate. 

\begin{figure}
\centering
\resizebox{0.99\hsize}{!}{
\includegraphics[clip=true,trim=0 0 0 0]{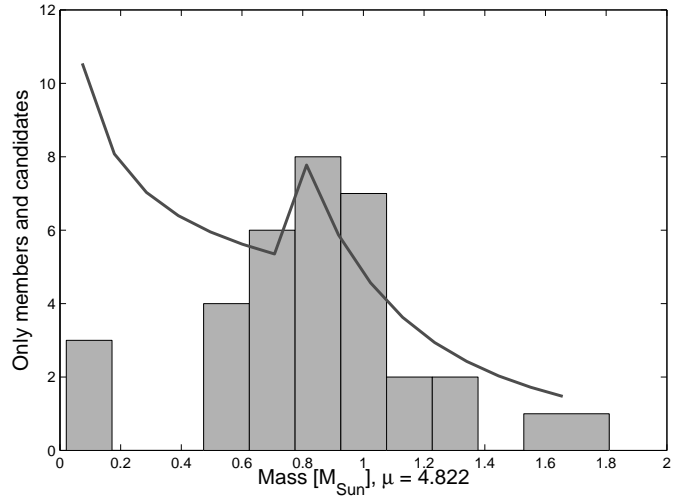}
}
\caption{Mass function of the Her-Lyr association compared to the initial mass function (IMF). Only good candidates (membership criterion + and ? in Table \ref{memtab}) are included. The number of stars is $N=\mu M^{-\alpha}$ with the stellar mass $M$ in $\Msun$, the proportionality factor $\mu$, and the index $\alpha$ (see text). Only stars with $M\geq0.8\Msun$ are used to fit $\mu$. See text for further explanations.
}
\label{imf}
\end{figure}

The mass function of the Her-Lyr association, when compared to the initial mass function, suffers from various selection effects and should therefore be used with care (Fig.~\ref{imf}). The low mass content of the Her-Lyr association is dramatically underestimated in the MF since the census of  K, M, and brown dwarfs is not at all complete. Thus, the low mass members are only represented through the companions, found close to various Her-Lyr members. Observations predict that the number of stars is given by 
\[
N=\mu M^{-\alpha},
\]
with the proportionality factor $\mu$ and the index 

\begin{equation}
\alpha=\left \{
\begin{array}{lclrlclcl}
2.35 & \mathrm{for} & &&M&>&1.0&\Msun &\mbox{\citep{1955ApJ...121..161S}}\\
2.3   &  \mathrm{for} & 0.5&<&M&\leq&1.0&\Msun &\mbox{\citep{2002Sci...295...82K}}\\
1.3  & \mathrm{for} & 0.08&<&M&\leq&0.5&\Msun &\mbox{\citep{2002Sci...295...82K}}\\
0.3  & \mathrm{for} & &&M&\leq&0.08&\Msun &\mbox{\citep{2002Sci...295...82K}}.\\
\end{array} 
\right .
\end{equation}

In Fig.~\ref{imf}, the MF is binned the same way as the histogram and the free parameter $\mu$ is adjusted to the star count above $0.8\,\Msun$, to use only bright stars for the fit which are counted properly. Figure~\ref{imf} suggests that the stellar content of Her-Lyr is complete down to $\sim 0.6\,\Msun$. The general trend of the MF is well reproduced above $\sim 0.6\,\Msun$. 

\section{Outlook}
\label{outlook}
In the past, the { very} existence of the Her-Lyr association was doubted by various authors \citep[e.g.,][]{2010A&A...521A..12M}. At the very least, following \citet{2004AN....325....3F}, the Her-Lyr association is an individual entity in kinematic space. Applying our selection criteria, the often-discussed heterogeneity in age and evolutionary state vanished to some extent but, a large age spread is still present. It is not clear if this is due to the uncertainties of the method these ages were derived with, or if this is an intrinsic property of the stars considered as members. However, the existence of the Her-Lyr association, as its own entity, or at least as a sub-moving-group of the LA is becoming clearer.

The Her-Lyr association displays a substantial spread in { gyrochronological} age and 
lithium absorption. If one assumes that this spread is real and 
intrinsic to Her-Lyr, there are important implications for the nature of 
this association. Although, a spread in lithium is well-known for 
the Pleiades which have similar age, there could also be a 
sub-structure in Her-Lyr which has not been resolved yet because of the small number of known members.

The Her-Lyr association could be part of the \LA which 
is a mixture of Pleiades-like stars in the solar neighborhood \citep{1956AJ.....61..405E}. The 
unclear composition of the \LA is reflected by the 
uncertainty of further sub-structure in Her-Lyr. The X-ray luminosity function, which is similar to that of the Pleiades, supports this idea. Future work clearly 
needs to address the relationship of Her-Lyr with other local moving 
groups (e.g., AB Dor, \citealp{2004ApJ...613L..65Z}, SACY, \citealp{2006A&A...460..695T}, etc.) and to 
show in detail whether the local moving groups are different from Her-Lyr.

Recently, a self-consistent study of the UMa and Hyades clusters revealed that the UMa association could be only about $100\,$Myr younger than the Hyades (\citealp[][but also see]{2005PASP..117..911K} \citealp[][]{2004AN....325....3F}), i.e., significantly older than the Her-Lyr association. 
The Hercules-Lyra association could therefore be an intermediate case in between the Pleiades (which are $100$ to $110\,$Myr old, \citealp[see][respectively]{1993A&AS...98..477M,2000AJ....119.1303T}) and UMa. This impression is strengthened by the comparison of lithium equivalent width (Fig.~\ref{lifig}) and by the average { gyrochronological} age of about $250\,$Myr derived for Her-Lyr. However, the spread in lithium of Her-Lyr is generally comparable to that of the Pleiades and seems to be larger than in UMa.
 In the future, lithium abundance needs to be studied instead of lithium equivalent widths. { Stellar parameters, as well as iron and lithium abundance,} have to be studied in a self-consistent way \citep[cf.][]{2009A&A...508..677A}.

For many candidates subject to the present work, spectroscopic 
information or rotational periods are missing. Additional observations are 
needed to obtain the missing data for these stars and to update the 
member list and criteria.

The distribution of non-members in the $UV$-space (see Fig.~\ref{hluvp}) possibly corresponds to a $\sim 1\,$Gyr old stellar stream. This has to be investigated in the future.

The mass function (Fig.~\ref{imf}) is well investigated down to $\approx 0.7\,\Msun$. This is further evidence that there is a large number of unidentified low mass stars in the solar neighborhood, as was pointed out by \citet{2004AN....325....3F} and other studies before.

Hercules-Lyra is an interesting laboratory for astrophysics. Similar to the 
Hyades, UMa group, and other local moving groups, the members are bright 
so that even late-type members can be studied in detail. The Her-Lyr association 
provides an important additional snapshot in the rotational and Li 
evolution of young stars. The stars are close so that companions can be 
resolved and common proper motion can be assessed quickly. Low-mass 
companions are particularly interesting since they possibly have not yet 
reached the main-sequence or might even be brown dwarfs at a very early 
cooling stage. { The multiplicity study presented in this work contributes to the total stellar census of young moving groups.}

\begin{acknowledgements}
The authors want to thank the staff of ESO Paranal and of the-ppl Calar Alto observatory, for helping to carry out observations used in this work. 
TE would like to thank the  {\it Deutsche Forschungsgemeinschaft} (DFG), projects  SFB/TR-7 and NE 515/30-1
for financial support. M.A. acknowledges research funding granted by the DFG under the project RE 1664/4-1. M.A. further acknowledges support by {\it Deutes Zentrum f\"ur Luft- und Raumfahrt} (DLR) under the projects 50OO1007 and 50OW0204. T.O.B.S. and R.N. like to thank the DFG under project NE 515/30-1. A.B. and R.N. would like to thank the DFG for financial support in projects NE 515/13-1 and 13-2. T.R., C.A., and R.N. acknowledge support from DFG project NE 515/23-1 and NE 515/33-1 in SPP 1385: "The first ten million years of the solar system". TE thanks L. Trepl and M. Corcoran for help with the ROSAT data. We thank the anonymous referee and the A\&A editors for inspiring suggestions and their patience.
This research has made use of the SIMBAD and VizieR database, operated at CDS, Strasbourg, France, and NASA Astrophysics Data and NASA's Astrophysics Data System.
and of data and software provided by the High Energy Astrophysics Science Archive Research Center (HEASARC), which is a service of the Astrophysics Science Division at NASA/GSFC and the High Energy Astrophysics Division of the Smithsonian Astrophysical Observatory. Last but not least we would like to thank the anonymous referee, who made numerous suggestions to improve the article and even suggested additional analysis methods to secure our results.
\end{acknowledgements}
\bibliographystyle{aa}
\bibliography{hlplit}
\end{document}